# Ferroionic states in ferroelectric thin films

*Anna N. Morozovska,*[1,2] *Eugene A. Eliseev,*[1,3] *Nicholas V. Morozovsky*[1]*, and Sergei V. Kalinin*[4,*]

[1]Institute of Physics, National Academy of Sciences of Ukraine,
46, Prospekt. Nauky, 03028 Kyiv, Ukraine

[2]Bogolyubov Institute for Theoretical Physics, National Academy of Sciences of Ukraine,
14-b Metrolohichna str. 03680Kyiv, Ukraine

[3]Institute for Problems of Materials Science, National Academy of Sciences of Ukraine,
3, Krjijanovskogo, 03142 Kyiv, Ukraine

[4]Center for Nanophase Materials Science, Oak Ridge National Laboratory, Oak Ridge, Tennessee 37831, USA

**Abstract**

The electric coupling between surface ions and bulk ferroelectricity gives rise to a continuum of mixed states in ferroelectric thin films, exquisitely sensitive to temperature and external factors, such as applied voltage and oxygen pressure. Here we develop the comprehensive analytical description of these coupled ferroelectric and ionic ("ferroionic") states by combining the Ginzburg-Landau-Devonshire description of the ferroelectric properties of the film with Langmuir adsorption model for the electrochemical reaction at the film surface. We explore the thermodynamic and kinetic characteristics of the ferroionic states as a function of temperature, film thickness, and external electric potential. These studies provide a new insight into mesoscopic properties of ferroelectric thin films, whose surface is exposed to chemical environment as screening charges supplier.

[*] Corresponding author. E-mail: sergei2@ornl.gov  (S.V.K.)

# I. INTRODUCTION

Ferroelectrics as primary ferroics have remained one of the traditional focal points for condensed matter physics research since the discovery of ferroelectricity in inorganic perovskites in 40-ies [1, 2, 3]. This interest was driven by the applications ranging from SONAR, optoelectronic systems, to recently non-volatile memories, ferroic heterostructures, micro- and nanoelectromechanical systems [4, 5, 6, 7]. In parallel, the basic physical understanding of ferroelectrics was developed, from bulk phenomenological to atomistic theories to explore finer aspects of polarization dynamics, order parameter couplings, and properties of topological defects [8, 9, 10, 11, 12]. Since the early days of ferroelectricity, it was recognized that key aspects of ferroelectric physics are the effects induced by the polarization bound charges that appear in the region with polarization discontinuities or gradients, including surfaces, interfaces, and topological defects [13, 14, 15]. Generally, the assumption that bound charges are uncompensated leads to major problems in thermodynamic or atomistic descriptions of ferroelectric state, since even at the macroscopic level the free energy of the system with constant polarization diverges with size, and on the level of Landau-Ginsburg-Devonshire (**LGD**) theory the ferroelectric phase is completely suppressed by depolarization fields [10, 14, 16, 17]. Correspondingly, much of the earlier analysis assumed that polarization charges are screened in some *ad hoc* manner and their effect can be ignored [18].

However, this assumption can no longer be used for description of ferroelectric surfaces and interfaces, where the interplay between polarization and screening charges defines the physical behavior of the system. At interfaces and topological defects, the screening is realized via classical semiconducting band-bending or redistribution of mobile ionic species [10, 14, 19]. The specific cases of this behaviors for linear or non-linear ferroelectric or rigid semiconductors have been explored by a number of authors [20, 21, 22, 23, 24, 25, 26]. In several recent studies, the role of more complex coupling mechanisms such as flexoelectricity was studied as well [27, 28, 29, 30]. Often, the interplay between linear and non-linear screening gives rise to complex phase diagrams. It is also important to note that much of the recent interest to polarization screening at interfaces and topological defects was induced by advances in instrumental techniques such as Scanning Transmission Electron Microscopy (STEM) [31, 32, 33], Piezoresponse Force Microscopy (PFM) [34, 35, 36, 37, 38, 39] and conductive Atomic Force Microscopy (cAFM) [40, 41, 42, 43], that allowed to probe functional properties of individual defects and probe the atomic structure, order parameter fields, and local oxidation states with atomic resolution. More generally, interfaces and topological defects offer an advantage of being relatively well defined – the observed phenomena are confined by the atomic lattice of the materials, for which experimental observations are possible and mesoscopic and atomistic models are well defined (unlike e.g. grain boundaries, where segregation composition of the cores represents major, and difficult to access, element of the model).

However, surfaces of ferroelectric materials present significantly more complex problem. Similarly to interfaces, surfaces necessitate the presence of the screening charge to compensate the ferroelectric bound charge [13]. Moreover, the surface state of materials in contact with atmosphere is usually poorly defined, due to the presence of mobile electrochemically active and physisorbable components in ambient environment, the issue well known in surface science community [44, 45]. The effect of adsorption of oxygen and hydrogen on the work function, spontaneous polarization value and switching, and also dipole moment of the unit cell and free energy of the semiconductor ferroelectrics was investigated experimentally and theoretically [46, 47, 48, 49]; however, the total research effort was miniscule compared to more classical aspects of ferroelectric behaviour and no comprehensive and/or general models were developed.

Surface screening of spontaneous polarization in ferroelectrics are typically realized by the mobile charges adsorbed from the ambient in the case of high or normal humidity [50, 51, 52, 53, 54]. In a specific case of the very weak screening, or its artificial absence due to the experimental conditions (clean surface in inert atmosphere or ultra-high vacuum) the screening charges can be localized at surface states induced by the strong band-bending by depolarization field [13, 55, 56, 57, 58]. For both aforementioned cases the screening charges are at least partially free (mobile), in the sense that the spatial distribution of their quasi two-dimensional density is highly sensitive to the changes in polarization distribution near the surface. When the screening charges follow the polarization changes almost immediately, the screening density can be calculated theoretically in the adiabatic approximation [13, 58]. In the opposite case, one should solve dynamic relaxation type equations for the spatial-temporal distributions of screening charges [58]. Due the long-range nature of the depolarization effects, the incomplete surface screening of ferroelectric polarization strongly influences the domain structure and leads to its pronounced changes both near and relatively far from the surface [59]. As such, it directly affects such phenomena as domain wall pinning at surfaces and interfaces, nucleation dynamics, domain shape and period control in thin film under the open-circuited conditions [9, 60], phase evolution in films between imperfect "real" electrodes with finite Tomas-Fermi screening length [61] or separated from the electrodes by ultra-thin dead layers [62] or physical gaps [63], emergence of closure domains near free surfaces in multiaxial ferroelectrics [9, 64, 65], domain wall broadening in both uniaxial and multiaxial ferroelectrics [66, 67] and crossover between different screening regimes of the moving domain wall - surface junctions [68, 69].

The aforementioned references mostly consider the physical aspects of the linear screening of spontaneous polarization in ferroelectrics, for which the screening charge density in the electrodes is linearly proportional to the surface bound charge (namely to its average over the film thickness). However, these treatments, while exploring the properties of ferroelectric material in detail, typically ignored the nonlinear tunable characteristics of surface screening charges. Recently, a complementary thermodynamic approach was developed by Stephenson and Highland (**SH**) [70, 71], who consider an ultrathin film in equilibrium with a chemical environment that supplies ionic species (positive and negative) to compensate its polarization bound charge at the surface. Equations of state and free energy expressions of a single-domain ferroelectric film are based on LGD theory, using electrochemical equilibria to provide ion compensating boundary conditions. The film top surface was exposed to a controlled oxygen partial pressure in equilibrium with electron conducting bottom electrode. The stability and metastability boundaries of phases of different polarization orientations ("positive" or "negative") are determined as a function of temperature and oxygen partial pressure in dependence on the film thickness.

Here we develop the analytical approach to explore the static and dynamic properties of coupled surface ionic and ferroelectric states in thin ferroelectric films, further referred to as **"ferroionic"** states. We extend the thermodynamic SH approach for description of the dynamic coupling between the ferroelectric polarization and surface ions. Due to the presence of the gap between the top electrode (e.g. SPM tip) and ferroelectric surface the surface charge density is controlled by the internal potential that depends on the applied electric voltage in a rather complex way. The internal potential is very similar to the **overpotential** commonly appearing in electrochemical problems.

The manuscript is structured as following. The mathematical statement of the problem is given in **section II**. **Section III** explores the peculiarities of ferroelectric polarization and surface charge coupling, characteristic length-scales of size effects and establishes the stability conditions of the single-domain state with out-of-plane polarization. The impact of the surface ion charge controlled by Langmuir adsorption on the (over)potential, average internal electric field and polarization dependences on the applied voltage is studied in **section IV**. The screening impact on the polarization, phase diagrams of ferroelectric thin films and their size effect is analyzed in **sections V and VI.** In **Section VII** the analysis of the surface screening impact on polarization reversal kinetics is presented. Distinctive features of polarization screening by ions are summarized in **section VIII**. The derivation of the equations coupling the dynamics of polarization and surface ion charges, polarization relaxation and its reversal kinetics calculations in LGD approach are given in **Supp. Mat**.[72].

**II. PROBLEM STATEMENT**

The time-dependent LGD formalism for the description of polarization dynamics in ferroelectric film in the case of SPM geometry is presented in **subsection A**. SH approach modified for the description of dynamical screening of polarization bound charges by sluggish ion charges located at the ferroelectric film surface is presented in **subsection B**. Analytical formalism for the description of polarization screening by electron conducting electrodes, and under the gap presence is discussed in **subsections C** and **D.**

**A. Ferroelectric polarization dynamics within Landau-Ginzburg-Devonshire approach**

To understand the coupling between ferroelectric and electrochemical phenomena, we consider the uniaxial ferroelectric film, for which the polarization components $P_i$ depend on the inner field **E** as $P_1 = \varepsilon_0(\varepsilon_{11}^f - 1)E_1$, $P_2 = \varepsilon_0(\varepsilon_{11}^f - 1)E_2$ and $P_3(\mathbf{r}, E_3) = P_3^f(\mathbf{r}, E_3) + \varepsilon_0(\varepsilon_{33}^b - 1)E_3$, where a relative background permittivity $\varepsilon_{ij}^b \leq 10$ is introduced [22]. The dielectric permittivity $\varepsilon_{33}^f$ related with the ferroelectric polarization $P_3$ via the soft mode is typically much higher than that of the background, $\varepsilon_{33}^f \gg \varepsilon_{33}^b$. The evolution and spatial distribution of the ferroelectric polarization $P_3$ is given by the time-dependent LGD equation [16]:

$$\Gamma \frac{\partial P_3}{\partial t} + a_3 P_3 + a_{33} P_3^3 + a_{333} P_3^5 - g_{33} \frac{\partial^2 P_3}{\partial z^2} = E_3, \quad (1a)$$

with the boundary conditions (**BCs**) of the third kind at the film surfaces $z = 0$ and $z = h$ [73]:

$$\left(P_3 - \Lambda_- \frac{\partial P_3}{\partial z}\right)\bigg|_{z=0} = 0. \quad \left(P_3 + \Lambda_+ \frac{\partial P_3}{\partial z}\right)\bigg|_{z=h} = 0. \quad (1b)$$

The physical range of extrapolation lengths $\Lambda_+$ and $\Lambda_-$ are (0.5 – 2) nm [74]. The film thickness is $h$ [**Fig. 1**].

Kinetic coefficient $\Gamma$ is defined by phonon relaxation and corresponding Landau-Khalatnikov (**LK**) time $\tau_K = \Gamma/|a_3|$ is small (~($10^{-11} - 10^{-13}$) s) far from immediate vicinity of the ferroelectric phase transition [75]. The critical slowing down occurs in the vicinity of the phase transition [3], at that relaxation times from 1 μs to 1 ms correspond to vicinity of transition temperature from several to several tens degrees, and hence this case is not considered here. Coefficient $a_3 = \alpha_T(T - T_C)$ is temperature dependent, $T$ is the absolute temperature, $T_C$ is Curie temperature. Coefficients $a_{33}$, and $a_{333}$ are the coefficients of LGD potential expansion on the polarization powers. In thin strained films, coefficient $a_3$ is renormalized by electrostrictive coupling [76, 77]. Quasi-static electric field is defined via electric potential as $E_3 = -\partial \varphi_f / \partial x_3$, where the potential $\varphi_f$ satisfies a conventional electrostatic equation inside the ferroelectric film (see below).

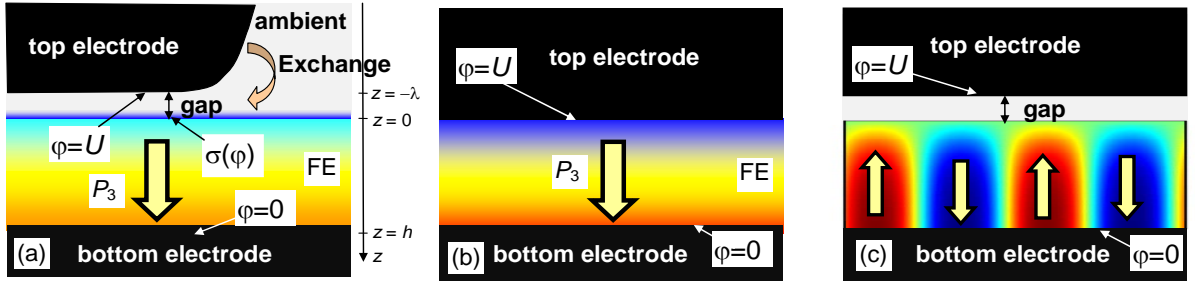

**FIGURE 1**. **(a)** Layout of the considered system, consisting of electron conducting bottom electrode, ferroelectric (FE) film, surface ions with charge density $\sigma(\varphi)$, ultra-thin gap separated film surface and the top electrode (either ion conductive planar electrode or flatted apex of SPM tip) providing direct ion exchange with ambient media (from bottom to the top). **(b-c)** Single-domain ferroelectric film placed in the ideal electric contact with electron conducting electrodes **(b)**; and the gap is present between the multi-domain film and the top electrode **(c)**. Figures **(c)** and **(b)** demonstrate the case with and without domains, respectively.

**B. Polarization screening by surface ions**

Here, we assume that the ion layer with the Langmuir-type isotherm density [70, 71] cover the ferroelectric film surface $z = 0$, and so the surface charges can screen the film polarization under the presence of the gap between the top electrode (or flat apex of the SPM tip) and the film surface [**Fig. 1(a)**]. We further consider several possible combinations of boundary conditions differing in the presence of dielectric layer and availability of the screening charges (electrochemical control vs. closed boundary conditions). We note that experimentally these can be realized either by design (similar to electrochemical experiments), or serendipitously via adsorption from environment.

For considered ultra-thin gap of thickness $\lambda$ (see **Fig. 1(a)**) the linear equation of state $\mathbf{D} = \varepsilon_0 \varepsilon_d \mathbf{E}$ that relates the electrical displacement $\mathbf{D}$ and electric field $\mathbf{E}$ is valid. Here $\varepsilon_0$ is a universal dielectric constant, and $\varepsilon_d$ is the relative permittivity of the physical gap media (vacuum, air or inert gas environment), so $\varepsilon_d \approx 1$. The bottom electrode is regarded flat and ideally electron conducting.

The system of electrostatic equations in each of the medium (gap and ferroelectric film) acquires the form:

$$\Delta \varphi_d = 0, \qquad \text{(inside the gap } -\lambda \leq z \leq 0\text{)} \qquad (2a)$$

$$\left(\varepsilon_{33}^{b}\frac{\partial^{2}}{\partial z^{2}}+\varepsilon_{11}^{f}\Delta_{\perp}\right)\varphi_{f}=\frac{1}{\varepsilon_{0}}\frac{\partial P_{3}^{f}}{\partial z}, \quad \text{(inside the ferroelectric film } 0<z<h\text{)} \quad (2b)$$

3D-Laplace operator is $\Delta$, 2D-Laplace operator is $\Delta_{\perp}$.

Boundary conditions to the system (2) are **(i)** the equivalence of the electric potential to the applied voltage $U$ at the top electrode surface $z=-\lambda$ (or SPM tip apex modeled by the flat region); **(ii)** the equivalence of the normal component of electric displacements to the surface charge density $\sigma(\varphi)$ at $z=0$; **(iii)** the continuity of the electric potential and normal component of displacements $D_3=\varepsilon_0 E_3+P_3$ and $D_3=\varepsilon_0\varepsilon_d E_3$ at gap - ferroelectric interface $z=0$; and **(iv)** zero potential at the bottom electron conducting electrode surface $z=h$. Hence the boundary conditions have the form:

$$\varphi_d\big|_{z=-\lambda}=U, \quad (\varphi_d-\varphi_f)\big|_{z=0}=0, \quad \varphi_f\big|_{z=h}=0, \quad (3a)$$

$$\left(\varepsilon_0\varepsilon_d\frac{\partial\varphi_d}{\partial z}+P_3^{f}-\varepsilon_0\varepsilon_{33}^{b}\frac{\partial\varphi_f}{\partial z}-\sigma\right)\bigg|_{z=0}=0. \quad (3b)$$

To describe the coupling between ferroelectric phenomena and interfacial electrochemistry (redox reactions), we adopt the SH approach [70, 71] and further extend it by allowing for the presence of the dielectric gap between the top electrode (e.g. tip apex) and ferroelectric surface. For the case the voltage $U$ is applied to top electrode, and so the electric potential $\varphi$ is not equal to $U$ at the interface $z = 0$ due to the presence of the dielectric gap.

The equilibrium surface charge density $\sigma_0[\varphi]$ at the interface $z = 0$ is controlled by the relative concentration of surface ions $\theta_i$

$$\sigma_0[\varphi]=\sum_i\frac{eZ_i\theta_i(\varphi)}{A_i}, \quad (4a)$$

where $e$ is an elementary charge, $Z_i$ is the charge of the surface ions/vacancies, $A_i$ is the minimal area per ion at saturation (surface steric limit), at that $i=2$ to reach the charge compensation. The surface charge is controlled by the potential $\varphi$ at the interface $z=0$ in a self-consistent manner:

$$\theta_i(\varphi)=\frac{\alpha_i(\varphi)}{1+\alpha_i(\varphi)}, \quad \alpha_i(\varphi)=\left(\frac{p_{O2}}{p_{atm}}\right)^{1/n_i}\exp\left(\frac{-\Delta G_i^{00}-eZ_i\varphi}{k_B T}\right). \quad (4b)$$

Here $p_{O2}$ is the oxygen partial pressure excess (relative to atmospheric pressure $p_{atm}$), $n_i$ is the number of surface ions created per oxygen molecule, $\Delta G_i^{00}$ is the free energy of the surface ion formation at $p_{O2}=1$ bar and $U=0$. Equation (4) is analogous to the Langmuir adsorption isotherm used in interfacial electrochemistry for adsorption of neutral species onto a conducting electrode exposed to ions in a solution [70, 71, 78]. Equations (4) can be incorporated in the numerical or phase-field formalism [79]. We further note that the developed solutions are insensitive to the specific details of the charge compensation process, and are sensitive only to the thermodynamic parameters of corresponding reactions.

To describe the surface charge dynamics, we adopt a relaxation equation

$$\tau \frac{\partial \sigma}{\partial t} + \sigma = \sigma_0[\varphi], \qquad (5)$$

where the dependence of equilibrium charge density on electric potential $\sigma_0[\varphi]$ is given by Eqs.(4). We further note that surface screening necessitates motion of ionic species (rather than dipole reorientation), and may require long-range transport of neutral species and their dissociation or ionization. Hence the characteristic time of the surface charge relaxation $\tau$ can be expected to be much higher than the polarization relaxation time $\tau_K$. Thus, the charge dynamics (coupled to polarization via the potential) is expected to control the system relaxation and hence directly affects the phenomena such as domain wall motion, switching, etc. Note that in general the relationship between the relaxation rate and the state of the system can be complex [69, 80]; however, Eq. (5) provides the first order approximation valid close to the thermodynamic minimum.

**C. Polarization screening by electron conducting electrodes**

For comparison with the case of surface screening by ions, we consider a uniaxial ferroelectric film placed in ideal electric contact with perfect electron conducting electrodes [$\lambda = 0$, see **Fig. 1(b)**], further named as the **"ideal screening"** case. Here, only electrostatic equation (2b) should be solved with the boundary conditions $\varphi_d\big|_{z=0} = U$ and $\varphi_f\big|_{z=h} = 0$. Domain formation is not favorable in the case, since the emergence of domain walls only increases the positive gradient energy and does not lead to any decrease of the depolarization field energy. The result agrees well with previous studies [see e.g. Ref.[9] and refs therein].

For this case, the approximate analytical expressions for the dependences of spontaneous polarization and thermodynamic coercive field on the temperature and film thickness have been derived and shown to be valid with the high degree of accuracy (see [16] and [72]). The critical thickness of the film at which paraelectric phase becomes unstable has the form [16]:

$$h_{cr}^0(T) \approx \frac{g_{33}}{\alpha_T(T_C - T)}\left(\frac{1}{\Lambda_+ + L_C} + \frac{1}{\Lambda_- + L_C}\right), \qquad (6)$$

where $L_C = \sqrt{\varepsilon_0 \varepsilon_{33}^b g_{33}}$ is the longitudinal correlation length [81]. Note that the critical thickness (6) diverges at Curie temperature ($h_{cr}^0(T_C) \to \infty$ due to $T_C - T$ vanishing), while for the case of natural boundary conditions ($\Lambda_\pm \to \infty$) it tends to zero. Note that electrostriction and flexoelectric coupling can renormalize the gradient coefficient and extrapolation lengths in Eq.(6) (see refs. [82, 83]). The flexoelectric coupling should contribute to the response, but its relative contribution is assumed to be small.

**D. Polarization screening by electrodes in the presence of dielectric gap**

More complex case is the film placed between electron conducting metal electrodes, without any screening charges ($\sigma = 0$) at the film surfaces, but in the presence of **dielectric gap** ($\lambda > 0$) [see **Fig. 1(c)**]. Domain stripes formation can be favorable above the critical thickness of the film transition in a paraelectric phase because the domain walls appearance leads to the noticeable decrease of the positive energy of depolarization field [see e.g. Refs.[9, 62, 84] and refs therein]. If the gap thickness is greater than (0.05 – 0.5) nm the noticeable domain wall surface broadening occurs at the open surface of the film because it decreases the depolarization field energy. For the same reason the closure domains can appear in unstrained multiaxial ferroelectric films because the polarization rotation is possible in the case, at that their properties appeared rather sensitive to the flexoelectric coupling [85]. The domains disappear when the film thickness decrease below the critical value, at which it becomes paraelectric. The critical thickness is significantly smaller than the thickness of the single-domain state instability that can be estimated from the expression Eq.(S1) in Ref.[72].

**III. COUPLING OF POLARIZATION AND SURFACE CHARGE**

In this section, we derive the coupled equations for the average polarization and surface charge [**subsection A**], discuss the characteristic length scales of size effects [**subsection B**], describe relevant physical variables and their numerical values [**subsection C**], study the surface charge conditioned by Langmuir adsorption of ions [**subsection D**] and establish the stability conditions of the single-domain state with out-of-plane polarization towards formation of multi-domain state or paraelectric phase [**subsection E**].

**A. Coupled equations for the average polarization and surface charge**

Since the stabilization of single-domain polarization in ultrathin perovskite films takes place due to the chemical switching (see e.g. [70, 71, 86, 87, 88]), we a priory consider a **single-domain** film covered with ions, which charge is given by Eq.(4) [see **Fig. 1(a)**], and assume that its polarization distribution is sufficiently smooth. Posteriori we will briefly discuss the conditions of the single-domain state stability with respect to the domain formation.

In order to develop approximate analytical description of coupled ferroelectric-ionic system, we consider the one-dimensional case and assume that distribution is close to homogeneous, $P_3 \approx \langle P_3 \rangle$, where polarization is averaged over film thickness. For ferroelectric film with a wide band-gap this assumption is valid for the case $\Lambda_\pm \gg \sqrt{\varepsilon_0 \varepsilon_{33}^b g}$, with the latter length-scale being very small (smaller than the lattice constant).

We note that the internal screening by space charge carriers can in principle change the polarization profile at surfaces and interfaces, corresponding to a prototypical ferroelectric-semiconductor model. However, the internal screening becomes essential for the carrier concentration more than $10^{26}\,\mathrm{m}^{-3}$ [13, 60], well above those expected for pristine or slightly doped perovskite $BaTiO_3$, $Pb(Zr,Ti)O_3$, $BiFeO_3$, $KNbO_3$ and pseudo-ilmenites $LiNbO_3$, $LiTaO_3$, etc [4, 13, 56, 63, 68]. We emphasize that in real materials with high internal conductivity the screening by surface ions and internal screening will compete, and defer the discussion of these cases for future studies.

After mathematical transformations described in **Appendix A** in Ref.[72] we derive two coupled nonlinear algebraic equations for the surface charge density $\sigma$ and polarization $\langle P_3 \rangle$ averaged over the film thickness:

$$\Gamma \frac{\partial \langle P_3 \rangle}{\partial t} + a_R \langle P_3 \rangle + a_{33} \langle P_3 \rangle^3 + a_{333} \langle P_3 \rangle^5 = E_{eff}(U, \sigma), \qquad (7a)$$

$$\tau \frac{\partial \sigma}{\partial t} + \sigma = \sigma_0 \left[ E_{eff}(U,\sigma) h - \frac{\lambda \langle P_3 \rangle h}{\varepsilon_0 (\varepsilon_d h + \lambda \varepsilon_{33}^b)} \right]. \qquad (7b)$$

The value

$$a_R = \alpha_T (T_C - T) + \frac{g_{33}}{h}\left(\frac{1}{\Lambda_+} + \frac{1}{\Lambda_-}\right) + \frac{\lambda}{\varepsilon_0(\varepsilon_d h + \lambda \varepsilon_{33}^b)}, \qquad (8a)$$

is the free energy coefficient $a_3$ renormalized by "intrinsic" gradient-correlation size effects (the term $\sim g_{33}$) and "extrinsic" depolarizing size effect (the term $\sim \lambda$) caused by a depolarizing field existence in the dielectric gap. Since both effects are additive in expression (8a), what gives the opportunity to study their impact separately. In the case $g_{33} = 0$ (or $\Lambda \to \infty$) the gradient-correlation effect is absent. Depolarizing size effect is absent without a dielectric gap ($\lambda = 0$).

The effective electric field introduced in Eq.(7b) is

$$E_{eff}(U,\sigma) = \frac{\varepsilon_0 \varepsilon_d U + \lambda \sigma}{\varepsilon_0(\varepsilon_d h + \lambda \varepsilon_{33}^b)}. \qquad (8b)$$

Acting electric potentials linearly depends on the coordinate z as:

$$\varphi_d = U - \frac{z+\lambda}{\lambda}(U-\Psi), \qquad \varphi_f = (h-z)\frac{\Psi}{h}. \tag{9a}$$

Corresponding "effective" potential $\Psi$ that defines the equilibrium surface charge density is:

$$\Psi = h\left(E_{\mathit{eff}}(U,\sigma) - \frac{\lambda\langle P_3\rangle}{\varepsilon_0(\varepsilon_d h + \lambda\varepsilon_{33}^b)}\right). \tag{9b}$$

Expressions (7)-(9) couple ferroelectric polarization and surface ions of electrochemical nature.

The time dynamics of the screening charge given by equation Eq.(7b) acquires the form $\tau\frac{\partial\sigma}{\partial t}+\sigma=\sigma_0[\Psi]$, where the equilibrium density of the surface charge is determined from nonlinear transcendental equation,

$$\sigma(U)=\sigma_0[\Psi]\equiv\sigma_0\left[h\left(\frac{\lambda(\sigma-\langle P_3\rangle)+\varepsilon_0\varepsilon_d U}{\varepsilon_0(\varepsilon_d h+\lambda\varepsilon_{33}^b)}\right)\right] \tag{10}$$

**B. Characteristic length-scales of size effects**

Prior to deriving the analytical solution of Eq. (10), we analyze the relevant length scales of screening phenomena, which in turn should be compared with the film thickness $h$ and dielectric gap thickness $\lambda$. The first length scale is the intrinsic gradient length,

$$L_g = g_{33}\frac{(\Lambda_+ + \Lambda_-)}{\alpha_T T_C \Lambda_+ \Lambda_-}, \tag{11}$$

that defines the changes of polarization gradient near the film surfaces in the hypothetic case of the absence of the depolarization field (e.g. for the *a*-domain films with polarization parallel to the film surfaces). However, the depolarization field exists for the considered polarization orientation normal to the film surface (see **Figs.1**). From the last term in Eq.(8a) the "bare" depolarization length is

$$L_\lambda = \frac{\lambda}{\varepsilon_0\varepsilon_d\alpha_T T_C}. \tag{12}$$

Depolarization length $L_\lambda$ is fairly large and is expected to exceed 100 nm at $\lambda \geq 1$ nm. However, the length $L_\lambda$ does not control the extrinsic size effect in the presence of screening charge $\sigma$ at $z = 0$. Practically, $L_\lambda$ is the length scale of the depolarization field effect produced by the dielectric gap without surface screening, $\sigma = 0$.

In the presence of surface screening charges, $\sigma(\varphi)$, with relatively high minimal area per ion in saturation $A_i$ (and hence relatively small surface screening length $L_S$) the value of $\sigma$ is close to the full screening, $\sigma = \langle P_3 \rangle$ at $U = 0$. Assuming that $L_S$ is smaller than $L_\lambda$, it governs the size effect of the surface charge. Using the linear approximation in Eqs.(4) we derived the expression:

$$\sigma \approx \left( \langle P_3 \rangle - \varepsilon_0 U \frac{\varepsilon_d}{\lambda} \right) \left( 1 + L_S \left( \frac{\varepsilon_d}{\lambda} + \frac{\varepsilon_{33}^b}{h} \right) \right)^{-1} \tag{13}$$

Substituting Eqs.(13) and (8b) in Eq.(8a) we derived the approximate expression for the coefficient $a_R = \alpha_T T_C \left( \frac{T}{T_C} - 1 + \frac{1}{h}(L_g + L_D) \right)$. The condition $a_R = 0$ gives the critical thickness of the single-domain ferroelectric state disappearance:

$$h_{cr}(T) \approx \frac{T_C (L_g + L_D)}{(T_C - T)}, \tag{14}$$

where the length-scale of depolarizing size effect "dressed" by the screening charge is

$$L_D = \frac{L_S}{\varepsilon_0 \alpha_T T_C} \left( 1 + L_S \left( \frac{\varepsilon_d}{\lambda} + \frac{\varepsilon_{33}^b}{h} \right) \right)^{-1}. \tag{15}$$

The length $L_D$ can be much smaller that $L_\lambda$ under the condition $\lambda \gg L_S$, quite realistic due to the high concentration of mobile surface charges. One can rewrite the expression (14) for the critical temperature on the film thickness, namely $T_{cr}(h) \approx T_C \left( 1 - \frac{L_g + L_D}{h} \right)$.

Coupled Eqs.(7) allow to calculate the dependence of the surface charge on the electrical voltage, phase transitions temperatures and the average polarization depending on the thickness of the ferroelectric film and the dielectric gap, the oxygen partial pressure $p_{O2}$, the surface defects formation energy $\Delta G_i^{00}$, their maximal density $1/A_i$ and other parameters of the structure. Here, the numerical calculations were performed for structure **"bottom electrode - Pb$_{0.5}$Zr$_{0.5}$TiO$_3$ film - surface ions - gap - top electrode"**. **Table I** contains description of the main physical variables in Eqs.(1)-(15) and the numerical values of material parameters used.

**TABLE I. Description of physical variables and their numerical values**

| Description of main physical quantities used in Eqs.(1)-(16) | Designation and dimensionality | Numerical value for a structure Pb$_{0.5}$Zr$_{0.5}$TiO$_3$ film /surface ions/ gap/top electrode |
|---|---|---|
| Polarization of ferroelectric along polar axis Z | $P_3$ (C/m$^2$) | 0.75 for a bulk material |
| Coefficient of LGD functional | $a_3 = \alpha_T (T - T_C)$ (C$^{-2}$ J m) | T-dependent |
| Dielectric stiffness | $\alpha_T$ ($\times 10^5$ C$^{-2}$·J·m/K) | 2.66 * |

| Curie temperature | $T_C$ (K) | 665.75 * |
|---|---|---|
| Coefficient of LGD functional | $a_{33}$ ($\times 10^8$ J C$^{-4}$·m$^5$) | 1.91 * |
| Coefficient of LGD functional | $a_{333}$ ($\times 10^8$ J C$^{-6}$·m$^9$) | 8.02 |
| Gradient coefficient | $g_{33}$ ($\times 10^{-10}$ C$^{-2}$·J·m) | 1 |
| Landau-Khalatnikov relaxation time | $\tau_K$ (s) | $10^{-11} - 10^{-13}$ (far from Tc) |
| Thickness of ferroelectric layer | $h$ (nm) | 3 – 500 |
| Extrapolation lengths | $\Lambda_-$, $\Lambda_+$ (angstroms) | $\Lambda_-$=1 Å, $\Lambda_+$=2 Å |
| Surface charge relaxation time | $\tau$ (s) | >> phonon time |
| Thickness of dielectric gap | $\lambda$ (nm) | 0.4 |
| Permittivity of the dielectric gap | $\varepsilon_d$ (dimensionless) | 1 |
| Background permittivity of ferroelectric | $\varepsilon_{33}^b$ (dimensionless) | 10 |
| Universal dielectric constant | $\varepsilon_0$ (F/m) | $8.85 \times 10^{-12}$ |
| Electron charge | $e$ (C) | $1.6 \times 10^{-19}$ |
| Ionization degree of the surface ions | $Z_i$ (dimensionless) | $Z_1 = +2$, $Z_2 = -2$ |
| Number of surface ions created per oxygen molecule | $n_i$ (dimensionless) | $n_1 = 2$, $n_2 = -2$ |
| Minimal area per ion at saturation (surface steric limit) | $A_i$ (m$^2$) | $A_1 = A_2 = 2 \times 10^{-19}$ |
| Surface defect/ion formation energy | $\Delta G_i^{00}$ (eV) | $\Delta G_1^{00} = 1$, $\Delta G_2^{00} = 0.1$ |
| Gradient-correlation size effect lengthscale | $L_g = \dfrac{g_{33}(\Lambda_+ + \Lambda_-)}{\alpha_T T_C \Lambda_+ \Lambda_-}$ (nm) | 8.5 |
| Dielectric gap field lengthscale at $\sigma = 0$ | $L_\lambda = \dfrac{\lambda}{\varepsilon_0 \varepsilon_d \alpha_T T_C}$ (nm) | 255 (for $\lambda$ = 0.4 nm) |
| Depolarizing size effect lengthscale | $L_D \approx \dfrac{L_S}{\varepsilon_0 \alpha_T T_C}\left(1 + L_S \dfrac{\varepsilon_d}{\lambda}\right)^{-1}$ (nm) | 6.4 (for $L_S$ = 0.1 Å) |

* The coefficients $a_3$, $a_{33}$ and $a_{333}$ in Eq.(1), are related with coefficients $b_3$, $b_{33}$ and $b_{333}$ from Haun et al [89] as $a_3 = 2b_3$, $a_{33} = 4b_{33}$ and $a_{333} = 6b_{333}$, since the definition used by Haun et al for free energy [Eq.(1) in Ref.[89]] gives the following equation of state $2b_3 P_3 + 4b_{33} P_3^3 + 6b_{333} P_3^5 + ... = E_3$.

Note that the characteristic spatial scale $L_g$ of the gradient-correlation size effects is about 8.5 nm. The scale is of same order than the "dressed" length of depolarization size effects $L_D = 6.4$ nm for Pb$_{0.5}$Zr$_{0.5}$TiO$_3$ parameters and $L_S$ = 0.1 Å. Both scales are nearly two orders of magnitude less than the "bare" length $L_\lambda = 255$ nm. Therefore, the influence of size effects can be ignored for the films with a thickness more than 100 nm.

**D. Surface charge conditioned by Langmuir adsorption**

The dependence of the charge density $\sigma_0(\varphi)$ on electric potential φ was calculated from Eqs.(1)-(4) for the different values of oxygen partial pressure (varying from $10^{-6}$ to $10^6$ bar). These dependences are shown in **Figs. 2(a)-(c)**. Two steps are clearly seen. They originate from the two Langmuir adsorption isotherms given by Eqs.(4) and correspond to positive and negative charges. As one can see the step position is slightly pressure-dependent and strongly dependent on the surface ion species (ion or vacancy) formation energies $\Delta G_i^{00}$. Since here we have chosen $A_1 = A_2$ and $n_1 = -n_2 = 2$, the charge $\sigma_0$ is zero at zero potential $\varphi = 0$, $p_{O2} = p_{atm}$ 1 bar, and $\Delta G_1^{00} = \Delta G_2^{00}$ [**Fig.2(a)**], while $\sigma_0$ is nonzero at $\varphi = 0$ for different formation energies $\left(\Delta G_1^{00} - \Delta G_2^{00}\right) \geq 0.9$ eV [**Fig. 2(b)-(c)**]. If the formation energies are different for positive and negative surface charges, due to their different nature, nonzero $\sigma_0$ at $\varphi = 0$ originated from a nonzero interfacial potential $\varphi \neq 0$. The diffusion of $\sigma_0$-steps increases with the temperature increase because of thermal factor $k_B T$ in Eqs.(4b) [**Fig. 2(c)**]. For a given parameters the steps occur in the vicinity of potential values $\varphi \approx 0$ [**Fig. 2(a)**] and $\varphi \approx -0.5$ V [**Fig. 2(b)-(c)**], which correspond to the almost complete depletion of positive surface charge and accumulation of negative one respectively.

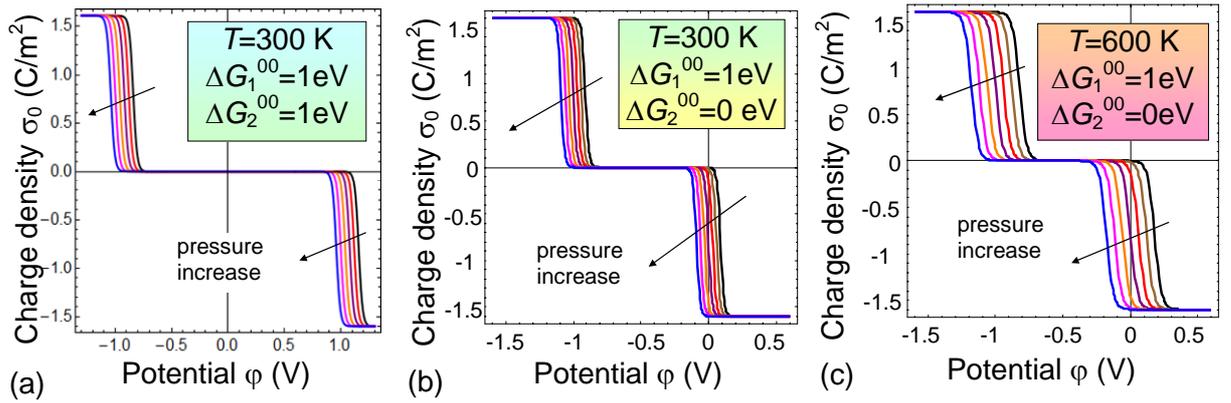

**FIGURE 2**. Dependence of the equilibrium surface charge density $\sigma_0$ vs. acting potential $\varphi$ calculated for the different values of oxygen partial pressure, $p_{O2} = 10^{-6}, 10^{-4}, 10^{-2}, 1, 10^2, 10^4, 10^6$ bar (black, brown, red, purple, orange, purple and blue curves respectively) at room temperature T = 300 K **(a-b)** and higher temperature T = 600 K **(c)**. Gap width $\lambda = 0.4$ nm. Ferroelectric parameters of $Pb_{0.5}Zr_{0.5}TiO_3$ and other parameters used in our calculations are listed in **Table I**.

**E. The stability conditions of a single-domain state with out-of-plane polarization**

We further briefly consider the instability of the ferroionic state towards domain formation. We note that in general such analysis requires consideration of free energy of the single domain ferroionic state as performed in this manuscript, and periodic domain state with surface chemical screening (since screening by definition lowers the free energy of the system) as a function of domain size. With these two energies, the tendency for domain formation and equilibrium domain size can be determined in a straightforward manner as a minimum of total energy. However, the full analysis of the thermodynamics of periodic domain state for prescribed electrochemical potential requires a separate and more involved study, beyond single domain case considered here. Hence, we perform elementary analysis of the domain instability and defer detailed studies to future work.

First, we consider the propensity for the in-plane domain formation. The coupled Eqs.(7) are derived for out-of-plane polarization direction. However, the in-plane polarization direction can be more stable than the out-of-plane one in the considered thin film of multiaxial ferroelectric $Pb_{0.5}Zr_{0.5}TiO_3$ under the absence of misfit strain or poling electric field [90, 91]. A compressive strain about -1 % should be applied to stabilize the out-of-plane polarization in the case of $Pb_{0.5}Zr_{0.5}TiO_3$ [90, 91]. However if the film is covered by the ion layer that creates its own electric field in the direction normal to the film surface, the field being strong enough can stabilize the out-of-plane polarization without any additional compressive strain. This situation is realized in our case [shown in **Fig. 1(a)**] if the difference between the ion formation energies $\left(\Delta G_1^{00} - \Delta G_2^{00}\right)$ is sufficiently high. Namely, the out-of-plane polarization direction is stable in thin $Pb_{0.5}Zr_{0.5}TiO_3$ films if $\left(\Delta G_1^{00} - \Delta G_2^{00}\right) \geq 0.3$ eV, because the screening ions create the strong polarizing electric field proportional to their charge $\sigma$, the "effective" value of which $E_{eff}(0,\sigma)$ is given by Eq.(8b). The field, being inversely proportional to the film thickness, is nonzero at zero applied voltage, $E_{eff}(0,\sigma) \neq 0$, because the ion charge density $\sigma(U)$ is nonzero at $U = 0$ for different ion formation energies $\Delta G_1^{00} \neq \Delta G_2^{00}$ (namely, we put $\Delta G_1^{00} = 1$ eV and $\Delta G_2^{00} \leq 0.1$ eV) and applied pressure more than 0.01 bar [see **Figs. 2(b-c)**].

Next we consider the system stability towards 180-degree domain formation. Note that even ultra-thin epitaxial film without top electrode (or with a gap between the film surface and the top electrode) can split into domain stripes [92, 93, 94]. However our analysis proves that the considered thin $Pb_{0.5}Zr_{0.5}TiO_3$ films are single-domain because the screening charge layer that covers the film surface creates the strong polarizing electric field, which effective value $E_{eff}(0,\sigma)$ is higher than coercive field of the film. Despite the ultra-thin gap between the sluggish ions and the top electrode is present in the geometry of this work, the situation with domain appearance seems completely different due to the strong polarizing role of the field at $U = 0$ and $\left(\Delta G_1^{00} - \Delta G_2^{00}\right) \geq 0.3$ eV. In the actual range of temperatures and thickness of $Pb_{0.5}Zr_{0.5}TiO_3$ films [shown in **Figs. 3(a)-(b)**] the field $E_{eff}(0,\sigma)$ becomes higher than the "intrinsic" thermodynamic coercive field $E_c$ corresponding to a homogeneous polarization reversal.

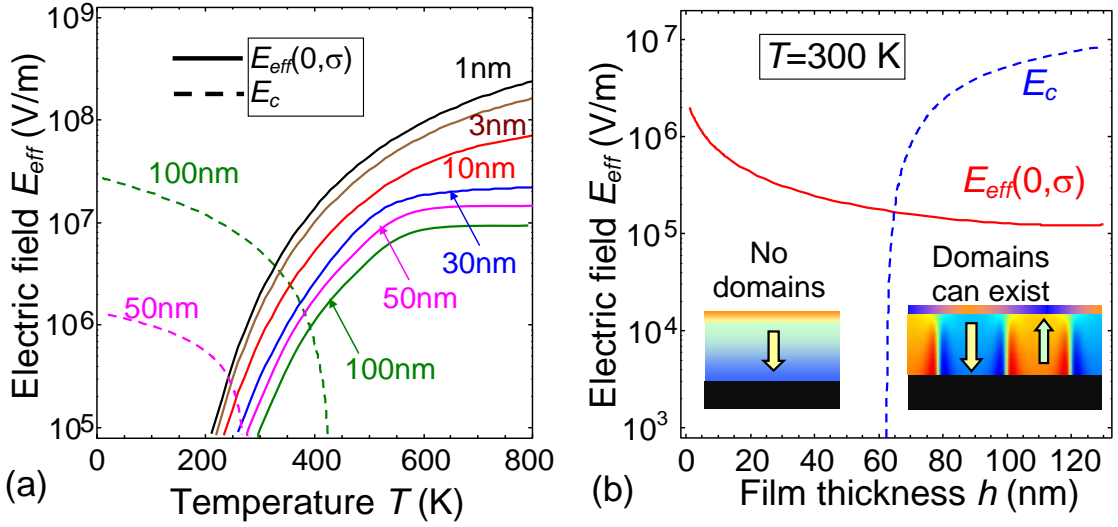

**FIGURE 3.** (a) Temperature dependences of the effective electric field $E_{eff}(0,\sigma)$ (solid curves) and thermodynamic coercive field $E_c$ (dashed curves) calculated for different thickness $h$ (from 1 to 100 nm) of $Pb_{0.5}Zr_{0.5}TiO_3$ film. (b) Thickness dependences of $E_{eff}(0,\sigma)$ (red curve) and $E_c$ (blue curve) calculated at room temperature. Parameters are described in **Table I**, $U = 0$.

The intrinsic coercive field was found numerically from Eqs. (7a). Its approximate analytical expression can be derived from Eqs. (11b) using the approach of Fridkin and Ducharme [10, 95], $E_c = \frac{2}{5}\left(2a_{33} + \sqrt{9a_{33}^2 - 20a_R a_{333}}\right)\left(\frac{2a_R}{-3a_{33} - \sqrt{9a_{33}^2 - 20a_R a_{333}}}\right)^{3/2}$, where the thickness-dependent coefficient $a_R$ is introduced in Eq.(8a). The coercive field $E_c$ is zero or negligibly small for the films of thickness less than 30 nm, because they are paraelectric under the absence of ion charge [dashed curves are absent for the thicknesses $h = (1 - 30)$ nm in **Fig. 3(a)**]. In these films the field caused by ions readily supports the single-domain state. The coercive field $E_c$ appears for the films thicker than 50 nm at temperatures less than 250 K, and becomes higher than $E_{eff}(0,\sigma)$ with the film thickness increase [dashed curves appear for the thicknesses $h = (50 - 100)$ nm in **Fig. 3(a)**]**.** Hence the films of thickness 50 nm and higher can split into ferroelectric domains, once the process becomes favourable (e.g. at $E_{eff}(0,\sigma) < E_c$). At the temperatures $T \geq 300$ K, the domains are absent for the film thickness less than 60 nm [see solid and dashed black curves in **Fig. 3(b)**].

The intrinsic coercive field $E_c$ corresponds to a homogeneous polarization reversal and thus gives strongly overestimated values in comparison with a realistic "extrinsic" coercive field. So that the field caused by ions indeed polarizes thin $Pb_{0.5}Zr_{0.5}TiO_3$ films and prevents the domain formation for the films of thickness at least smaller than $h_{SD} = 50$ nm for the gap thickness $\lambda = 0.4$ nm and arbitrary temperatures [**Fig. 3(a)**]. To obtain more realistic estimation of the threshold thickness $h_{SD}$ we treat the problem numerically [72] and obtained that the field of ion charge supports the single-domain state stability up to the thicknesses about (110-160) nm.

Based on this analysis, we argue that a (10 – 150)-nm $Pb_{0.5}Zr_{0.5}TiO_3$ film covered with ions remains single-domain with the out-of-plane polarization component $P_3$ for chosen boundary conditions. However in reality the thickness range of the single-domain state stability can be somewhat different due to the pinning and memory effects not accounted in our estimations.

**IV. EFFECTIVE FIELD AND OVERPOTENIAL DUE TO LANGMUIR ADSORPTION**

In describing the properties of the ferroionic state, we note that the potential φ at the interface z = 0 can be strongly different from the voltage $U$ applied to the electrode at $z = -\lambda$. In thermodynamic equilibrium, the surface ion charge σ located at the film-gap interface depends on the overpotential Ψ in accordance with Eq.(10). Exactly the charge controls the effective field $E_{eff}(U,\sigma)$ [see Eq.(8b)] and the average polarization $\langle P_3 \rangle$ dependence on applied voltage $U$ [see Eq.(7a)]. Thus it makes sense to analyze the analogous voltage dependences of the overpotential Ψ, ion charge σ, its field $E_{eff}$ and film polarization $\langle P_3 \rangle$.

The dependences of Ψ, σ, $E_{eff}$ and $\langle P_3 \rangle$ on applied voltage $U$ were calculated from the coupled Eqs. (7)-(8) and are shown in **Fig. 4**. Different curves correspond to different thicknesses of the film. Since we choose the significant difference $(\Delta G_1^{00} - \Delta G_2^{00}) = 0.9$ eV, the voltage dependences of the ion surface charge, effective field and polarization are strongly asymmetric for thin film with thickness $h < 50$ nm [see the curves for $h = 3, 10$ and $30$ nm in **Figs. 4**]. The asymmetry of Ψ(*U*) originates from the presence of surface charge. It is noticeable at $h < 50$ nm and becomes weak for thicker films, where the response of the ferroelectric film dominates [**Fig. 4(a)**]. The step-like dependences of Ψ(*U*) [**Fig. 4(a)**] and surface charge σ [**Fig. 4(b)**] on applied voltage $U$ cause the step-like voltage behavior of $E_{eff}$ [**Fig. 4(c)**] and $\langle P_3(U) \rangle$ [**Fig. 4(d)**]. The physical origin of $\langle P_3(U) \rangle$ behavior is the appearance of $E_{eff}(U,\sigma)$ that contains two terms in accordance with Eq.(7a). The first term (~*U*) is a linear drop of applied voltage due to the gap presence, the second term linearly depends on $\lambda\sigma(U)$ and manifests the step-like peculiarities, as well as it is nonzero at $U = 0$ if σ(0) is nonzero. The second term leads to the step-like peculiarities of $\langle P_3(U) \rangle$ dependence, and induces nonzero polarization at $U = 0$ for the films with thickness less than the critical one, $h_{cr} \cong 20$ nm [red curve for $h = 10$ nm in **Fig. 4(d)**]. The polar state at $h < h_{cr}$ mimics ferroelectric behavior due to the nonlinear dependence $\sigma(U)$.

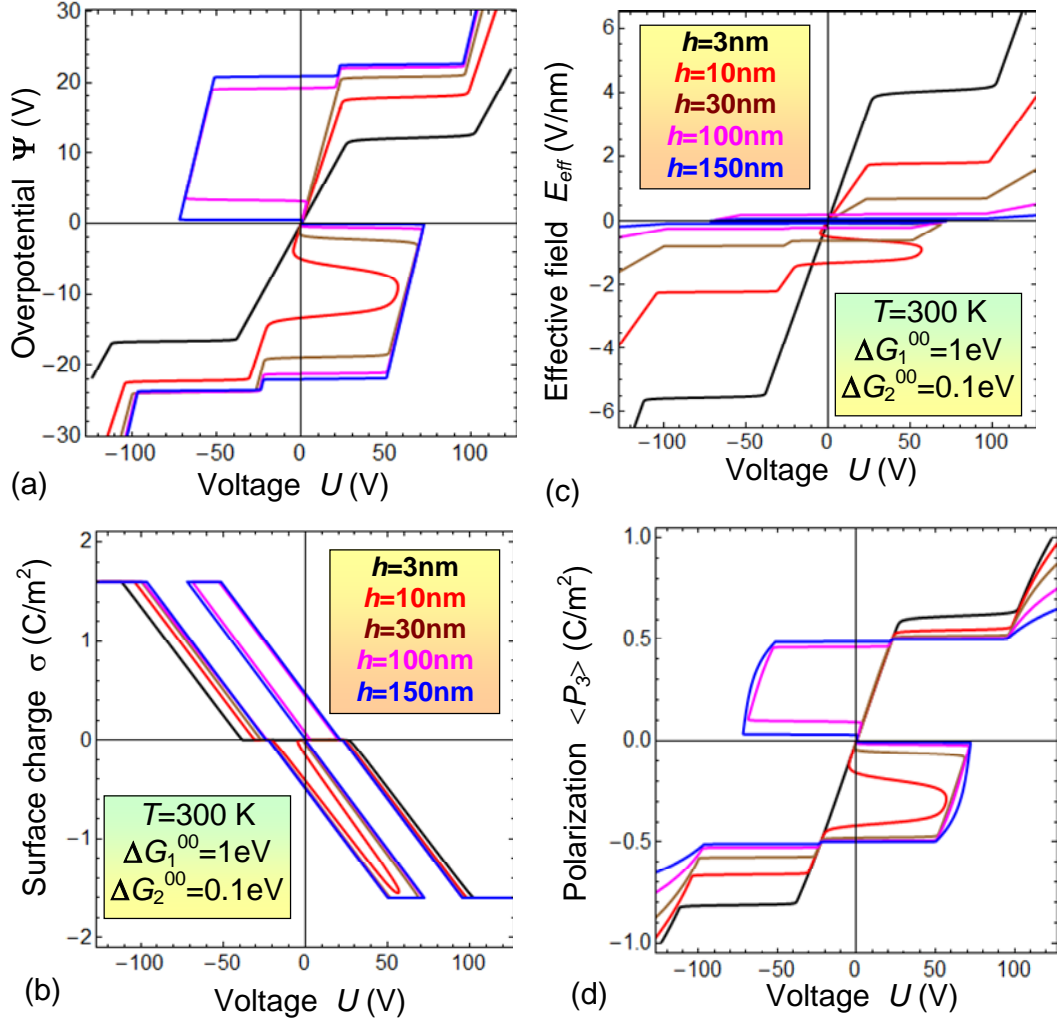

**FIGURE 4**. Dependences of the overpotential $\Psi$ **(a)**, surface charge $\sigma$ **(b)**, effective electric field $E_{eff}(U,\sigma)$ **(c),** and average polarization $\langle P_3 \rangle$ **(d)** on applied voltage U calculated for different thickness of ferroelectric film $h$ = (3, 10, 30, 100, 150) nm (black, red, brown, purple and blue curves). Parameters T = 300 K $p_{O2}$ = 1 bar, $\Delta G_1^{00} = 1$ eV, $\Delta G_2^{00} = 0.1$ eV, and $\lambda = 0.4$ nm. Other parameters used in our calculations are listed in **Table I**.

Note that in the case of ideal screening without any surface charges ($\lambda = 0$ and $\sigma = 0$) the overpotential coincides with the applied potential ($\Psi = U$), the effective electric field is as inside of the plane capacitor ($E_{eff} = U/h$). Corresponding voltage dependence the average polarization becomes a conventional S-characteristic for $h > h_{cr}^0$ ($h_{cr}^0 \cong 5$ nm) with much smaller coercive voltage that in the case of the gap and surface charge presence [see in **Fig. S2** in **Appendix B** in Ref.[72]].

To summarize, the unusual voltage dependence of the film polarization (step-like peculiarities and polar states in ultra-thin films) is caused by the ion-created poling electric field that effective value is given by Eq.(8a). The field destroys the transition from the ferroelectric to paraelectric phase at the critical thickness and the intermediate coupled ferroelectric and ionic (ferroionic) states appear instead. We further note that this yields unusual features of the hysteresis loops, films phase diagrams and polarization temperature dependences, as considered in the next sections.

**V. SCREENING EFFECT ON THE PHASE DIAGRAMS OF FERROELECTRIC THIN FILMS**

To establish the impact of polarization screening by adsorbed surface ions on the films phase diagrams, we calculated the diagrams for three physically important **cases 1-3** shown in **Fig.1(a)-(c)**, respectively.

The model **cases 1-2** with $\sigma = 0$ can be analyzed using the approach [16, 60]. To support the out-of-plane polarization direction, a compressively strained film (strain is about -1%) should be considered [90, 91]. The phase diagram of the film corresponding to the cases 1 and 2, calculated in the coordinates "temperature – film thickness" are shown in **Fig. 5 (a).** The critical thickness $h_{cr}$ of the film transition to a paraelectric phase is approximately equal to 5 nm at $\lambda = 0$ [**case 1**, the solid curve in **Fig. 5 (a)**], about 20 nm at $\lambda = 0.4$ nm allowing for the domain appearance [**case 2**, the dashed curve in **Fig. 5 (a)**]; at that the thickness $h_{SD}$ of the single-domain state instability is about 254 nm at $\lambda = 0.4$ nm [**case 2**, dotted curves in **Fig. 5 (a)**]. The phase diagram exhibits the classical ferroelectric behaviour [16], namely the film can be in the paraelectric (PE) or ferroelectric (FE) phases when its thickness is smaller or greater than the critical one, respectively. The boundary between PE and FE phases is sharp.

Finally, we consider the most interesting **case 3**, $\sigma \neq 0$. For correct comparison the gap thickness $\lambda$ was taken the same as in the **case 2** ($\lambda = 0.4$ nm). Corresponding phase diagrams were calculated in the coordinates "temperature – film thickness" and shown in **Figs. 5 (b-d)**. The oxygen partial pressure $p_{O2}$ varies in **Fig. 5(b)**, the surface ions saturation density $1/A_i$ changes in **Fig. 5(c)**, and the ion formation energy $\Delta G_2^{00}$ varies in **Fig. 5(d)** at fixed other parameters. As one can see, the phase diagram of the ferroelectric film screened by $\sigma \neq 0$ [**Figs. 5(b-d)**] differs dramatically from the diagram of the film with $\sigma = 0$ [**Fig. 5(a)**].

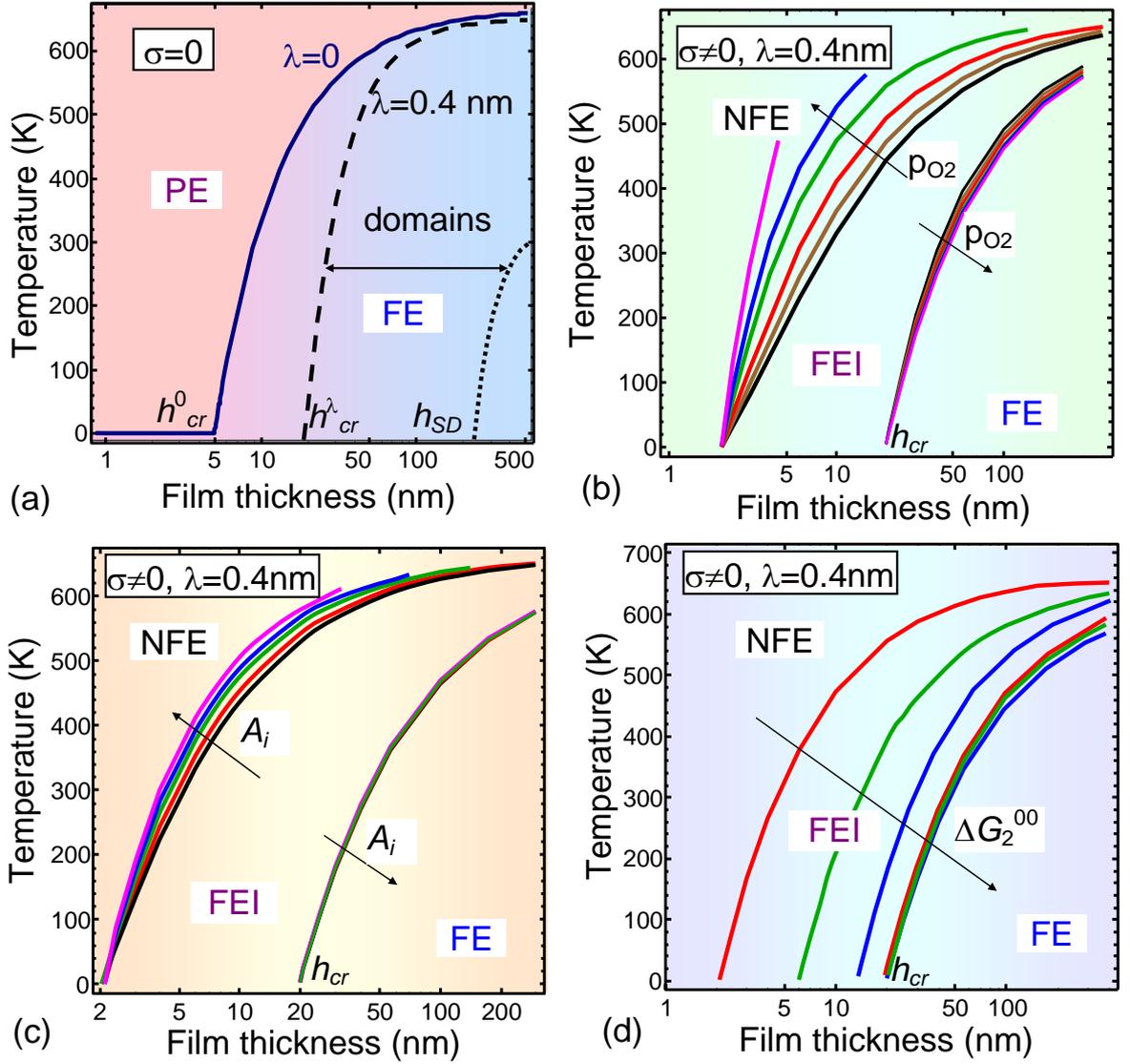

**FIGURE 5**. **(a)** Phase diagram in coordinates " $PbZr_{05}Ti_{05}O_3$ film thickness – temperature" calculated for the case of **ideal screening** of polarization by electron conducting electrodes (solid curve calculated for $\lambda = 0$) and under the presence of **dielectric gap** without any screening charges (dashed curve calculated for $\lambda = 0.4$ nm and $\sigma = 0$). Paraelectric (**PE**) and ferroelectric (**FE**) phases exist. **(b-d)** Phase diagram in coordinates "temperature – film thickness" calculated for the case of polarization screening by the **surface charge** with density $\sigma(\varphi)$ for different values of oxygen partial pressure $p_{O2}=$ ($10^{-7}$, $10^{-5}$, $10^{-3}$, $10^{-1}$, 1 and 10) bar (black, brown, red, green, blue and purple curves respectively) **(b)**; different values of the surface ions saturation density $N_s = 1/A_i = $ (2, 3, 5, 7 and 10)×$10^{18}$ m$^{-2}$ (black, red, green, blue and purple curves respectively) **(c)**; and different values of the ion formation energy $\Delta G_2^{00}=$(0.1, 0.3 and 0.5) eV (red, green and blue curves respectively) **(d)**. Arrows indicate the direction of $p_{O2}$ or $\Delta G_2^{00}$ increase, respectively. Non-ferroelectric ionic (**NFE**), ferroelectric-ionic (**FEI**) and ferroelectric (**FE**) states exist. Plot **(b):** $A_1 = A_2 = 2\times10^{-19}$ m$^{-2}$, $\Delta G_1^{00} = 1$ eV and $\Delta G_2^{00} = 0.1$ eV. Plot **(c):** $p_{O2} = 0.1$ bar,

$\Delta G_1^{00} = 1\,\text{eV}$, $\Delta G_2^{00} = 0.1\,\text{eV}$. Plot **(d):** $p_{O2}$ = 0.1 bar $\Delta G_1^{00} = 1$ eV and $N_s = 1/A_i = 5\times 10^{18}\text{m}^{-2}$. Other parameters are listed in **Table I**.

A general property of the diagrams shown in **Figs. 5(b-d)** is that without of applied voltage ($U=0$) the films can be in the non-ferroelectric ionic (**NFE**), ferroionic (**FEI**) and mostly ferroelectric (**FE**) states. Here, FE state is defined as the state with robust ferroelectric hysteresis between two absolutely stable and two unstable ferroelectric polarizations, which have "positive" or "negative" projection with respect to the film surface normal. The four polar states correspond to the four real roots of the static Eqs.(7a), $a_R \langle P_3 \rangle + a_{33} \langle P_3 \rangle^3 + a_{333} \langle P_3 \rangle^5 = E_{eff}(U,\sigma)$, which can exist at nonzero σ, film thickness $h \geq h_{cr}(T)$ and fixed temperature $T < T_C$. The positive and negative orientations of polarization are energetically equivalent only at σ = 0 and $U = 0$. The physical origin of the positive and negative polar states asymmetry is the field $E_{eff}(U,\sigma)$ produced by surface ions with charge density σ ≠ 0.

The FEI state appears once the film thickness $h$ becomes less than the critical one [see Eq.(15)], because positive polarization orientations loses their stability at $h \geq h_{cr}(T)$. The asymmetric hysteresis between two negative polarization orientations (one is the stable and another is the unstable root of the static Eq.(7a)) is supported by the ionic states bistability to applied voltage. Hence the boundary between FE and FEI states is exactly the dependence $T_{cr}(h)$. The relatively sharp boundary between FE and FEI states depends weakly on the oxygen pressure $p_{O2}$, surface ions saturation density $1/A_i$ or ion formation energy $\Delta G_2^{00}$ [see the very narrow bunches of colored curves in **Figs. 5(b-d)**].

Once two negative polarization orientations tend to split into the only one under the temperature increase (or further decrease of the film thickness well below the critical value), NFE state follows the FEI state. NFE state has no hysteresis properties in the thermodynamic limit, and can reveal electret-like polarization state $\langle P_3 \rangle \cong E_{eff}(\sigma)/a_R$ induced by the field $E_{eff}$. Since two negative polarization orientations gradually reach each other with the temperature increase the diffuse boundary between NFE and FEI states is defined by the electrochemical properties of surface ions, such parameters as the oxygen pressure $p_{O2}$, surface ions saturation density $1/A_i$, and ion formation energies $\Delta G_i^{00}$ [see the well-separated bunches of colored curves in **Figs. 5(b-d)**]. The curves divergence reflects the fact that NFE states are non-paraelectric, because the field $E_{eff}$ induces the irreversible "spontaneous" polarization of the film.

FEI states can exist even in ultrathin films ($h \leq 2$ nm) at high temperatures, depending on $p_{O2}$, $1/A_i$ and $\Delta G_2^{00}$ values.

Constitution diagrams in coordinates "partial oxygen pressure – temperature" calculated in the case of polarization screening by the ions are shown in **Figs 6.** Parts (a)-(d) corresponds to the film of different thickness $h = (20 - 150)$ nm.

Here, NFE, FEI and FE states exist, at that the temperature region of FE state stability noticeably increases with the film thickness. FEI state separates FE and NFE states. The boundary between FE and FEI states changes quasi-linearly with $p_{O2}$ for all thicknesses. The boundary between FE and FEI states changes with $p_{O2}$ in a super-linear fashion for all film thicknesses. Note the different temperature scales at **Figs. 6(a)-(b)** and **6(c)-(d)**. For the film thickness less than 30 nm FEI phase becomes stable at much lower temperatures (less than 200 K) than for $h \geq 100$ nm when the FE state loses its absolute stability above (450 - 550) K.

The non-trivial prediction of this analysis is the emergence of the critical point $S_{cr}$ at parameters $\{p_{cr}, T_{cr}\}$ terminating the boundary between NFE and FEI states. Note that the critical points in coordinates "partial oxygen pressure – temperature" have been revealed earlier by Highland et al. (see fig. 3 in Ref.[71]) and Stephenson et al (see fig. 17 in Ref.[70]). Following Landau [96], the phase equilibrium curve may end in some point that is called critical, and the corresponding temperature and pressure are the critical temperature and critical pressure. The critical point can exist only for such phases, the distinction between which is purely quantitative in nature. That say, the critical point is the inherent feature of the diagrams in coordinates "partial oxygen pressure – temperature" calculated in the case of polarization screening by the surface ions, which density is described by Langmuir functions (4) [see figure 12 in Ref. [70] and **Figs.2** in the work].

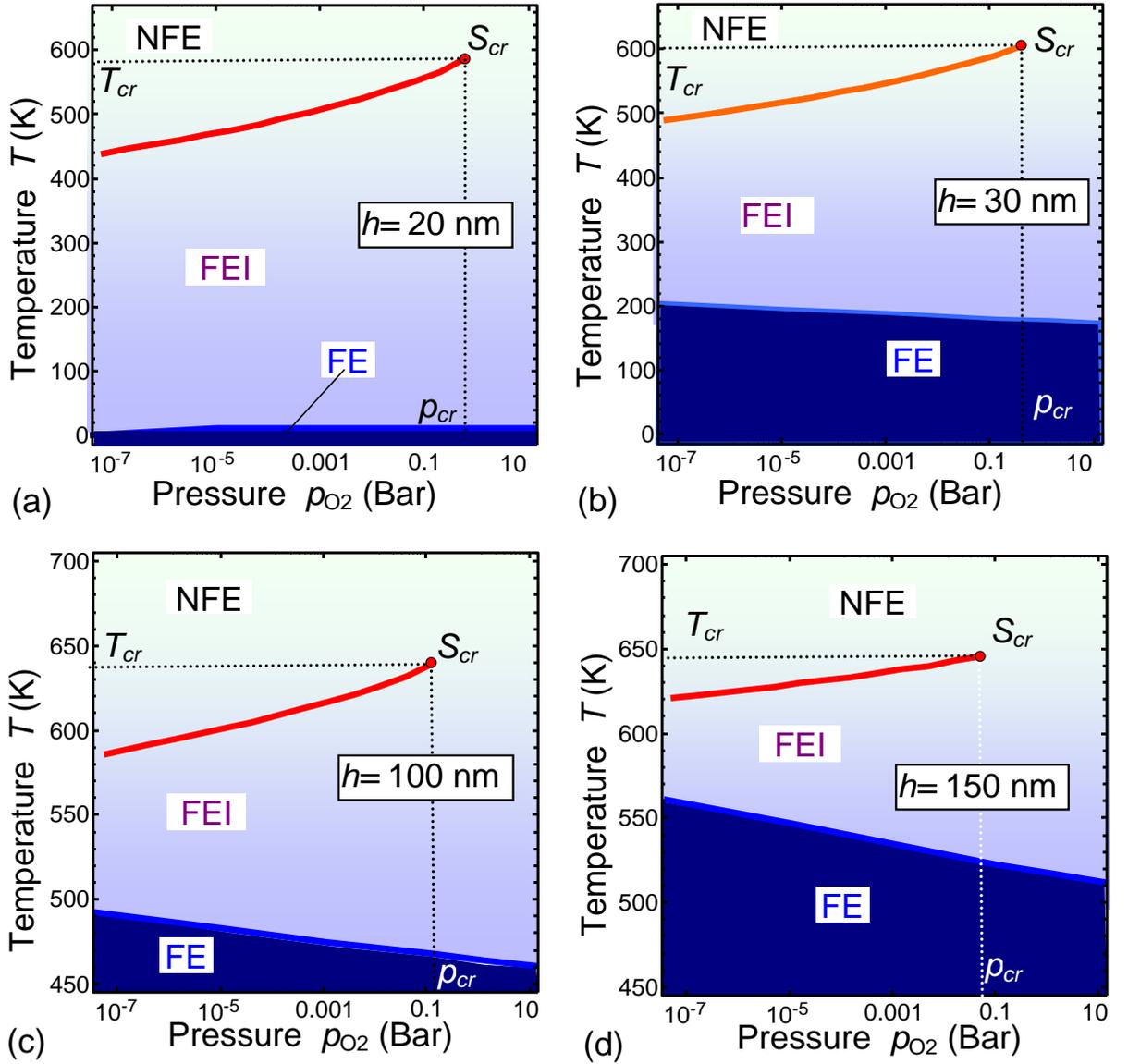

**FIGURE 6**. Constitution diagrams of $PbZr_{05}Ti_{05}O_3$ films in coordinates "partial oxygen pressure – temperature" calculated in the case of polarization screening by ions and different film thickness $h$ = 20 nm **(a)** and 30 nm **(b)**, 100 nm **(c)** and 150 nm **(d)**. $S_{cr}$ is the critical point. Non-ferroelectric ionic (**NFE**), coupled ferroelectric-ionic (**FEI**) and ferroelectric (**FE**) states exist. Material parameters are listed in **Table I**.

To **summarize**, thin ferroelectric film covered with ion layer that charge density $\sigma(\varphi)$ obeys Langmuir-type formulae (4) can be in NFE, FEI and FE states dependent on the film thickness, temperature and oxygen partial pressure $p_{O2}$. The critical thickness of the film transition into FE phase can be introduced. Sharp boundary between FEI and FE states weakly depends on $p_{O2}$. Diffuse boundary between NFE and FEI states noticeably depends on the pressure $p_{O2}$. The boundary between NFE and FEI states disappears in the critical point $S_{cr}$ that exists in the constitution diagram in coordinates "oxygen partial pressure $p_{O2}$ – temperature".

## VI. SCREENING IMPACT ON POLARIZATION THERMODYNAMICS

We further proceed to analyze the thermodynamics of the ferroionic states. Temperature dependences of the average polarization calculated for $\sigma \neq 0$ and $\sigma = 0$ (for comparison), and different values of film thickness $h = (3 - 150)$ nm are shown in **Figs. 7.**

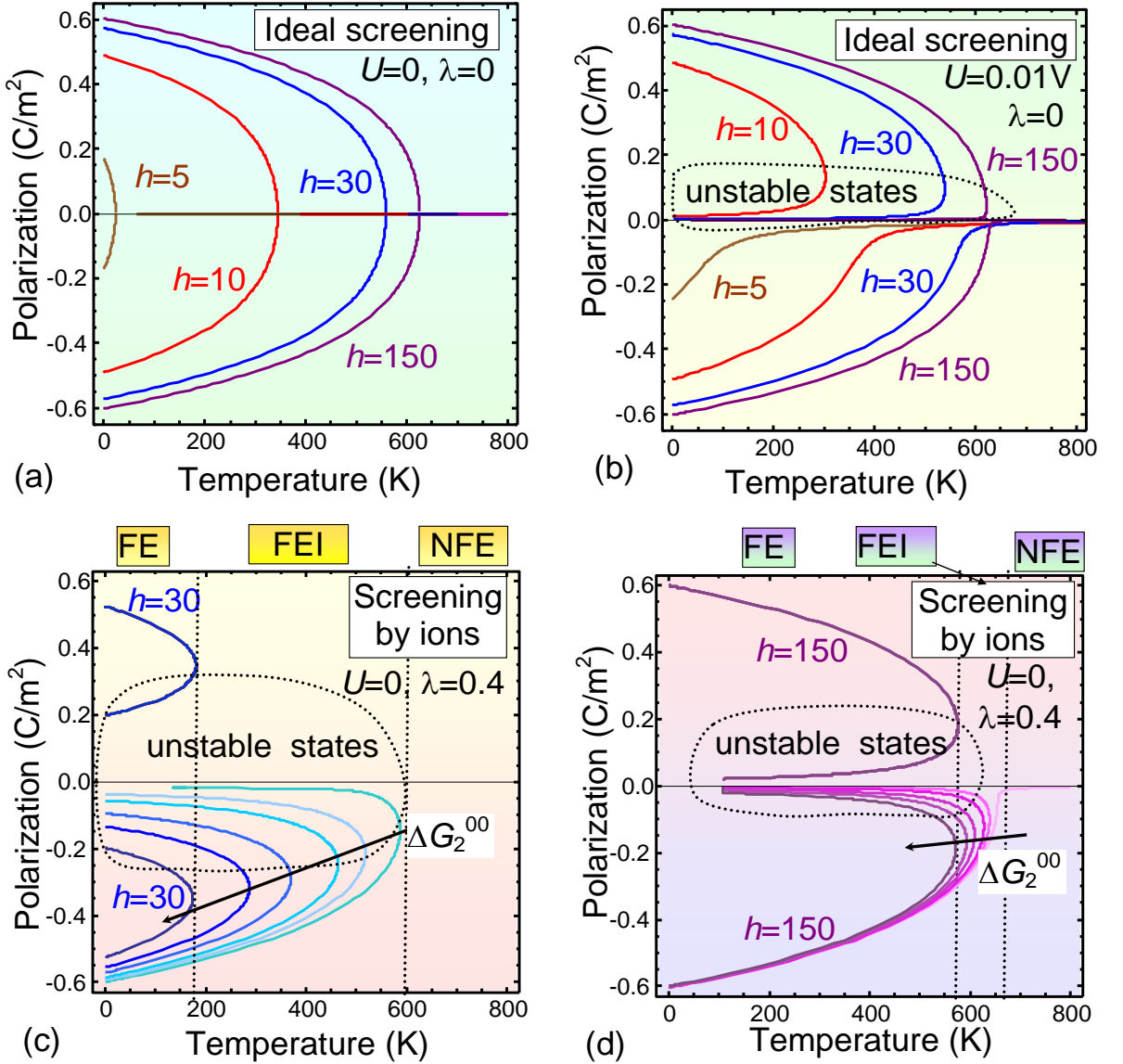

**FIGURE 7.** **(a)-(b)** Temperature dependences of average spontaneous polarization $\langle P_3 \rangle$ calculated in the case of polarization ideal screening by electron conducting electrodes ($\lambda = 0$ for plots **(a)** and **(b)**) without any screening charges ($\sigma = 0$), at zero **(a)** and nonzero **(b)** applied voltages U = 0, 0.01 V and 1 V, respectively. Different curves correspond to the different values of film thickness $h = (5 - 150)$ nm (legends near the curves). **(c)-(d)** Temperature dependences of the average polarization calculated in the case of **screening by ions**, zero applied voltage $U = 0$, $\lambda = 0.4$ nm, $A_1 = A_2 = 2 \times 10^{-19}$ m$^2$, $\Delta G_1^{00} = 1$ eV and different values of film thickness $h = 30$ **(c)** and 150 nm **(d)**. Each curve in the one-colored group

corresponds to the definite value of $\Delta G_2^{00} = (0, 0.1, 0.2, 0.3, 0.7)$ eV and 1 eV. Arrow indicates the direction of $\Delta G_2^{00}$ increase. Unstable (located inside the dotted area) and stable (located outside the dotted area) branches are shown in the plots (b) and (d). An arrow indicates the direction of $\Delta G_2^{00}$ increase. Material parameters are listed in **Table I**.

Curves in **Figs. 7(a)** and **7(b)** correspond to the case of polarization ideal screening ($\lambda = 0$) by electron conducting electrodes at zero ($U = 0$) and nonzero ($U = 0.01$ V) applied voltages respectively. Since both polarization directions are energetically equivalent at $U = 0$, the curves in **Fig. 7 (a)** are symmetrical with respect to the *T*-axis. Since Pb$_{0.5}$Zr$_{0.5}$TiO$_3$ is a ferroelectric with the second order phase transition, its spontaneous polarization occurs at the critical temperature $T_{FE}(h)$ and increases gradually with decreasing temperature, according to the relationship $P_3 \sim \sqrt{T_{FE}(h)-T}$. The temperature $T_{FE}(h)$ depends on the thicknesses of the film $h$ and gap $\lambda$, namely $T_{FE}(h)$ exists only for the $h$ values exceeding the critical one, $h_{cr} \approx 5$ nm (at $\lambda = 0$). For the thickness $h < h_{cr}$ the film is in the PE phase ($P = 0$ at $U = 0$), while the spontaneous polarization occurs for $h > h_{cr}$ and tends to the bulk value with the film thickness increase. The usual behavior of the spontaneous polarization and the transition temperature is analyzed in details in Refs. [14, 73, 97]. The curves lose their symmetry with respect to the polarization axis after applying even a small voltage to the top electrode [see **Fig. 7 (b)**], because the polarization direction for which $P_3 E_3 > 0$ is more favorable energetically. The probe field induces polarization of the negative sign, so the critical thickness disappears for negative polarization that asymptotically tends to zero with increasing temperature.

The characteristic temperature dependences of the average polarization calculated for the case of polarization screening by surface charges and zero applied voltage $U = 0$ are shown in **Figs. 7 (c)** and **7(d)** for 30-nm and 150-nm thick film, respectively. Since the screening by ion charges creates the internal poling field (Eq.(8b)), the curves in **Figs. 7(c)-(d)** are qualitatively similar to the corresponding curves in **Fig. 7(b)** calculated for a nonzero voltage $U$. Positive and negative polarization states become closer with the film thickness increase [compare **Fig. 7(c)** with **Fig. 7(d)**]. Each curve in the one-colored group corresponds to different values of ion formation energy $\Delta G_2^{00}$ at fixed other parameters. An arrow indicates the direction of $\Delta G_2^{00}$ increase. The curves corresponding to $\Delta G_2^{00}$ 1 eV are metastable states; they are shown for comparison only, because the domain formation and polarization rotation is favorable in PbZr$_{05}$Ti$_{05}$O$_3$ film in the case $\Delta G_1^{00} = \Delta G_2^{00} = 1$ eV. Note that the screening by ions is less effective than electron conducting electrodes for the positive polarization direction (compare upper parts of the **Figs. 7(a)** and **7(c,d)**) and more effective for the opposite negative direction [compare bottom parts of the **Fig. 7(b)** and **7(c,d)**].

We further analyze the temperature dependences of the polar states in the system in dependence on the electrochemical properties of surface screening charges. **Figures 8** show the temperature dependences of the average polarization calculated for the case at zero applied voltage. Different groups of blue, red and purple curves are calculated for film thickness 4 nm, 30 nm and 150 nm, respectively.

Every curve in the one-colored group corresponds to different values of oxygen partial pressure $p_{O2} = 10^{-7}$, $10^{-5}$, $10^{-3}$, $10^{-1}$, 1 and 10 bar in **Fig. 8(a)**. Each curve in the one-colored group corresponds to different values of ions saturation density $1/A_i = (2 - 10) \times 10^{18}$ m$^{-2}$ in **Fig. 8(b)**. Arrows indicate the direction of $p_{O2}$ and $1/A_i$ increase, respectively. Note that the range of $1/A_i$ values is chosen so that the corresponding charge densities $\sigma_S = e/A_i$ are comparable with the spontaneous polarization $P_S \approx 0.5$ C/m$^2$ of bulk PZT at room temperatures. In particular the highest value ($1/A_i = 10^{19}$ m$^{-2}$) corresponds to the maximal charge density $\sigma_S = 1.6$ C/m$^2$, more than three times higher than $P_S$. The minimal value ($1/A_i = 2 \times 10^{18}$ m$^{-2}$) corresponds to the density 0.32 C/m$^2$ that is about two times smaller than $P_S$. For the densities less than $2 \times 10^{18}$ m$^{-2}$ the FE states disappear for film thicknesses less than 30 nm. The behavior can be explained by the strong dependence of the critical thickness [Eq.(15)] on the surface charge density, because the surface screening length $L_S$ is inversely proportional to $A_i$. If the maximal density of screening ions is insufficient to screen the spontaneous polarization, the screening is performed by the top electrode separated from the film by the gap. In this case the critical thickness increases and the spontaneous polarization decreases in a self-consistent way according to Eqs.(7).

Comparing **Fig. 8(a)** and **Fig. 8(b)** we conclude that the pressure increase from $10^{-7}$ bar to 10 bar is equivalent qualitatively to the increase of ions saturation density. However for thin films the pressure changes affect the film phase diagram stronger than the changes of $1/A_i$ [compare the distance between the curves calculated for the same film thickness in **Fig. 8(a)** and **Fig. 8(b)**]. The comments made above regarding the range of $1/A_i$ are qualitatively suitable for the pressure changing, because the increase of $p_{O2}$ leads to the increase of the ions density, makes the screening more effective. However the concrete mechanism is different; the equilibrium surface charge density is linearly proportional to $1/A_i$ [see Eq.(4a)], and relative concentration of surface ions $\theta_i$ depend on $p_{O2}$ in a more complex way [see Eq.(4b)].

Since the screening by ion charges leads to appearance of poling electric field $E_{eff}(U,\sigma)$ inside the film (see Eq. (8b)), the curves in **Figs. 8** corresponding to the film thicknesses 30 nm and 150 nm, and relatively high oxygen pressures $p_{O2} \geq 1$ bar (or sufficiently big values $(1/A_i) \geq 2 \times 10^{18}$ m$^{-2}$) are qualitatively similar to the corresponding curves in **Fig. 7 (c)** and **7 (d)**, plotted for $U = 0.01$ V.

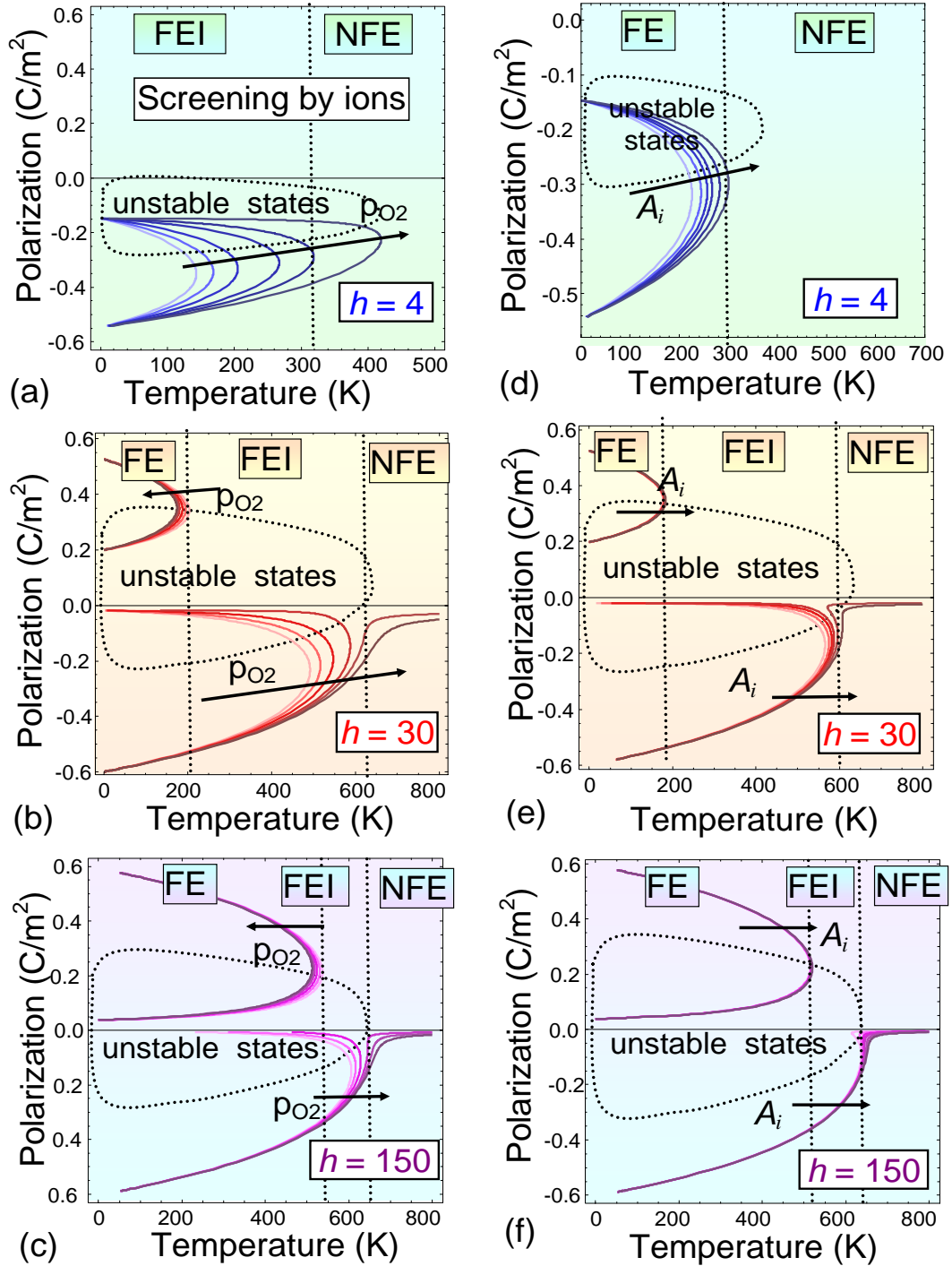

**FIGURE 8.** Temperature dependences of the average polarization $\langle P_3 \rangle$ calculated in the case of its screening by ions, zero applied voltage $U = 0$ and different values of PbZr$_{05}$Ti$_{05}$O$_3$ film thickness $h = 4$,

30 and 150 nm (groups of blue, red and purple curves in plots **(a)-(f)**). **(a)-(c)** Each curve in the one-colored group corresponds to the definite value of oxygen partial pressure $p_{O2}= 10^{-7}$, $10^{-5}$, $10^{-3}$, $10^{-1}$, 1 and 10 bar. Arrows indicate the direction of $p_{O2}$ increase; $A_1 = A_2 = 2\times 10^{-19}$ m$^2$, $\Delta G_1^{00} = 1$ eV and $\Delta G_2^{00} = 0.1$ eV. **(d)-(f)** Each curve in the one-colored group corresponds to the definite values of $1/A_i$ =(2, 3, 4, 5, 7 and 10)$\times 10^{18}$ m$^{-2}$. Arrows indicate the direction of $A_i$ increase; $p_{O2}$= 0.1 bar, $\Delta G_1^{00} = 1$ eV, $\Delta G_2^{00} = 0.1$ eV. Unstable (located inside the dotted area) and stable (located outside the dotted area) braches are shown. Material parameters are listed in **Table I**.

To **summarize** to section, under the absence of screening ion layer two equal spontaneous polarization orientations, +$P_S(T)$ and −$P_S(T)$, are stable and energetically equivalent at film thickness higher than the critical one ($h > h_{cr}$). The positive and negative spontaneous polarizations become not equivalent under the presence of surface ions. At that two branches of the negative spontaneous polarization exists and one of them is unstable at film thickness $h < h_{cr}$. At film thickness $h > h_{cr}$ four branches of the spontaneous polarization exists, two of them are negative and stable, and another two are positive and unstable. Negative polarizations are sensitive to the electrochemical characteristics of surface ions (oxygen pressure $p_{O2}$, their minimal area per ion $A_i$ and formation energies $\Delta G_i^{00}$).

**VII. SCREENING IMPACT ON POLARIZATION REVERSAL KINETICS**

Average polarization $\langle P_3 \rangle$ of the ferroelectric film, its time derivative $\partial \langle P_3 \rangle / \partial t$ (that is a displacement current for the case of gap presence and/or free carriers absence) and the charge density σ of surface ions dependences on the amplitude $U$ of the applied alternative voltage, $U_{\approx} = U \sin(\omega t)$, are analyzed in the section.

Prior to study the impact of the surface ions located at the gap-film interface on the dynamics of polarization based on the coupled Eqs.(7), which are derived in a single-domain approximation, let us analyze the system behavior against the domain appearance during the polarization reversal. The poling field $E_{\mathit{eff}}(U, \sigma)$ stabilizes the single-domain negative polarization at negative and zero applied voltages. Enough high positive voltage $U$ counteracts the field $E_{\mathit{eff}}(U, \sigma)$, and so the polarization reversal can be realized via both inhomogeneous or homogeneous reorientation scenario. To check both possibilities, we performed numerical simulations of polarization reversal in (10-100)-nm films at low (ωτ<<1) and intermediate (ωτ ~ 1) frequencies ω, which results demonstrated that the initial multi-domain seeding transforms into a single-domain state after a short transient process. After that the polarization reversal takes place via homogeneous scenario and the shape and sizes of the hysteresis loop calculated numerically are almost the same as calculated in the single-domain approximation from the coupled Eqs.(7). The inhomogeneous multi-domain switching is realized due to the strong retardation of the sluggish surface charge at high frequencies ($\omega\tau \gg 1$), which are not considered below.

In the case of ideal screening the shape and width of the polarization loops change substantially with increasing the dimensionless frequency $\omega\tau_K$ by an order of magnitude [the case of ideal screening is shown in **Figs. S4** of Ref.[72]]. The loop with finite width emerge even for the thinnest films, and the loops for thicker films become significantly wider and smoother. Corresponding current maxima become much lower and wider; the phase shift appears for the thickest film. At the frequencies $\omega\tau_K > 1$ all the loops become strongly "inflated". The conclusion is in a complete agreement with conventional Landau-Khalatnikov dynamics of homogeneous polarization reversal in perfectly screened ferroelectric films described by time-dependent LGD equation. Note that the gap presence increases the critical thickness of the loop appearance, leads to the loops smearing and causes the phase shift of the displacement current under the frequency increase [85].

To analyze the polarization reversal in the ferroelectric film covered by ions, we analyze the hierarchy of different relaxation times existing in the considered ferroionic system [**Fig.1(a)**]. The characteristic time $\tau$ of the surface charges relaxation was chosen to be much larger than the LK time $\tau_K \sim 10^{-11}$ s at room temperature (that is well below Curie temperature of the chosen ferroelectric material, $PbZr_{05}Ti_{05}O_3$). Numerically, the regime $\tau \gg 100\,\tau_K$ already corresponds to the adiabatic approximation, meaning that the polarization response to electric field is almost instantaneous in comparison with ion response. Since the further decrease of $\tau_K$ increases the time of calculation, and doest not add any new features to the results, below we analyze the case $\tau = 10^3\,\tau_K$.

In general case there is three different time scales, well below polarization relaxation, well below screening charges relaxation, and the intermediate range between them. Moreover, since we consider several types of the screening charges, ions, vacancies and electrons [see Eq.(4)], the electron (Maxwellian) relation time should be included into the hierarchy too. Taking into account the problem complexity for thin ferroelectric films with mobile electrons and ions [98], the impact of the relaxation times hierarchy on the polarization reversal in the presence of the surface charges will be studied elsewhere. Below we use the dimensionless frequency $\omega\tau$, but not the $\omega\tau_K$, because the LK time $\tau_K$ defines the timescale of polarization reversal kinetics only in the case $\sigma = 0$. Hence **Figs.9(a)** and **9(b)** were calculated for small ($\omega\tau =$ 0.1) and intermediate ($\omega\tau = 1$) dimensionless frequencies $\omega\tau$. Different colors of the curves correspond to different film thickness *h* changing in the range (3 – 100) nm**.**

The hysteresis loops of $\langle P_3(U)\rangle$ and $\partial\langle P_3\rangle/\partial t$ calculated at the lowest frequency $\omega\tau = 10^{-1}$ [shown in **Fig. 9 (a,c)**] are qualitatively similar to the corresponding hysteresis loops calculated in the case of polarization screening by ideal electrodes at $\omega\tau_K = 10^{-2}$ [shown in **Fig. S4(a,c)** of Ref.[72]]. A very narrow loop corresponds to the thinnest 3 nm thick film. The loop width increases with the film thickness increase and the shape tends to the square-like one at thickness 100 nm [see black, red, purple and blue loops in **Fig. 9(a)**]. Displacement currents have pronounced maxima at coercive voltage [see black, red, purple and blue loops in **Fig. 9(c)**]. However there are important quantitative differences between the loops shown in **Fig. 9** and LK dynamics in the case of polarization ideal screening. In particular, several minor features (thickening and jumps) are visible at polarization and current loops for large values of the applied voltage in **Fig.9 (a)** and **9(c)**. They occur due to the step-like voltage dependence of the surface charge [see **Fig. 10(a)**].

The behavior of the polarization and current hysteresis loop at intermediate ($\omega\tau =1$) frequencies seems paradoxical and cannot be reconciled with the LK dynamics in the case of polarization screening by ideal electrodes [compare **Figs.9 (b)** and **9 (d)** with **Figs.S4(b)** and **S4(d)**]. In particular the width of the loops decreases sharply and monotonically with the frequency increase, their shape becomes slim, and two cross-sections at the loop endings appear at the intermediate frequencies $\omega\tau =1$ [**Fig. 9 (b)**]. The sharp maxima of displacement current [**Fig.9 (c)**] is the contribution of the bound charge dynamics (ferroelectric or paraelectric polarization) in the film, and the wide maximum [**Figs.9 (d)**] is the retarded reaction of the surface charge on the polarization reversal in the film. The retardation appears because LK relaxation time is much smaller than the characteristic time of the surface charge relaxation.

The seemingly anomalous frequency dispersion of the loops shown in **Figs.9** can be easily rationalized. Here, the sluggish ion charges can screen the polarization at low frequencies ($\omega\tau \ll 1$) and strongly retard with respect to polarization reversal at higher frequencies ($\omega\tau \sim 1$)**.** Corresponding loops of the surface charge calculated in the frequency range $\omega\tau = (10^{-1} – 1)$ are shown in **Figs. 10(a)** and **10(b)**, respectively**.**

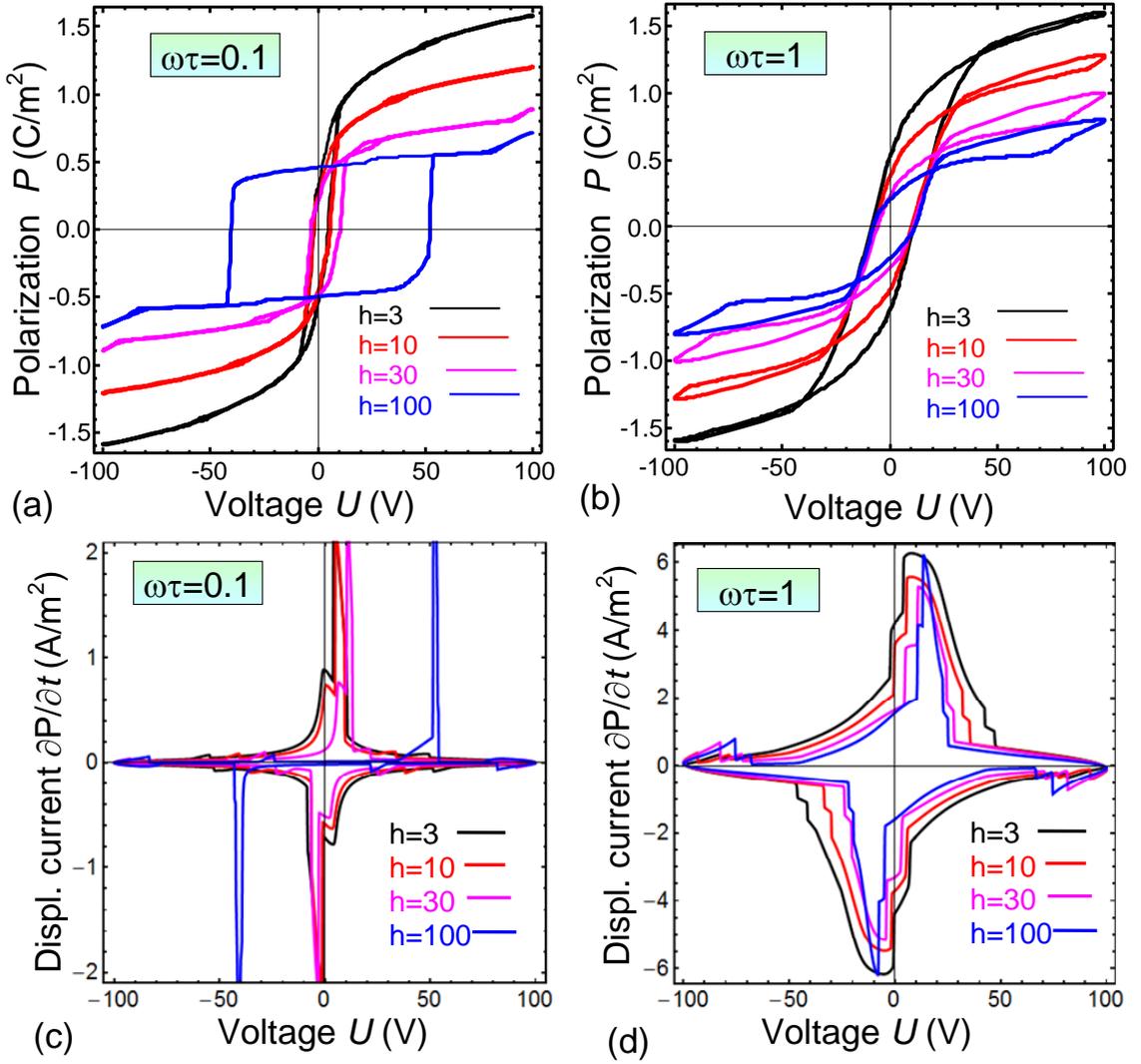

**FIGURE 9. Polarization reversal in the case of polarization screening by surface ions.** Dependences of the average polarization $\langle P_3 \rangle$ and displacement current $\partial \langle P_3 \rangle / \partial t$ on the amplitude $U$ of applied voltage calculated for its several frequencies $\omega$, $\omega\tau = 0.1$ **(a)** and $\omega\tau = 1$ **(b)**. Different colors of the curves correspond to different thickness of the film, $h = 3$ nm (black curves), 10 nm (red curves), 30 nm (purple curves) and 100 nm (blue curves). Temperature T = 300 K, oxygen partial pressure $p_{O2} = 10^{-2}$ bar, surface charge relaxation time $\tau = 10^3 \tau_K$. Other parameters used in calculations are listed in **Table I**.

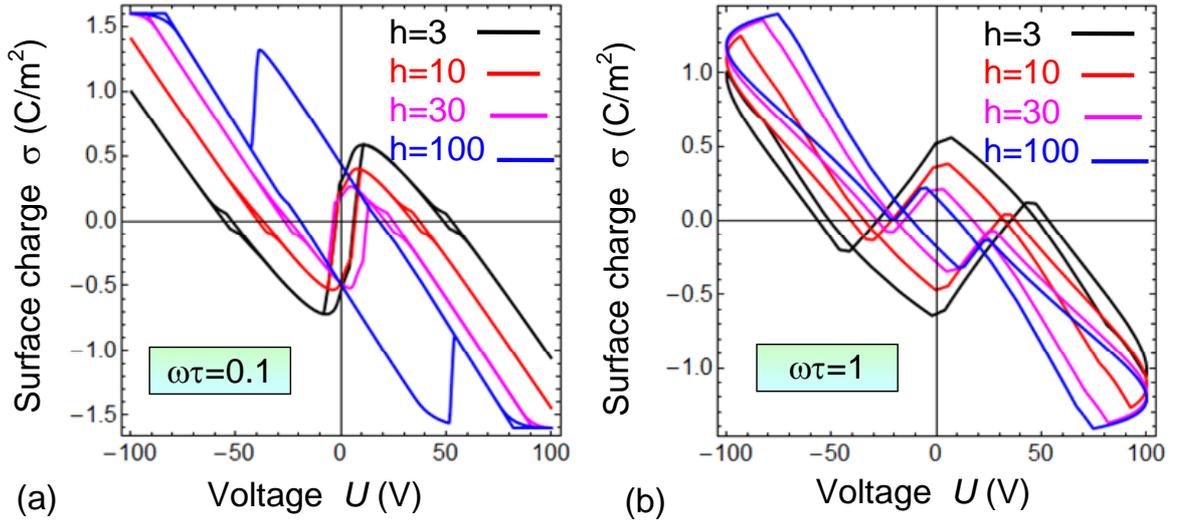

**FIGURE 10**. **Surface charge hysteresis.** Surface charge density dependence on the amplitude $U$ of applied alternative voltage calculated for its several frequencies $\omega$, $\omega\tau = 0.1$ **(a)** and $\omega\tau =1$ **(b)**. Different colors of the curves correspond to different thickness of the film, $h = 3$ nm (black curves), 10 nm (red curves), 30 nm (purple curves) and 100 nm (blue curves). Other parameters are the same as in **Fig. 9**.

Overall, we conclude that the revealed features of polarization reversal are a nontrivial manifestation of the screening charge dynamics in the framework of modified SH model. In particular, polarization hysteresis loops shape changes and their width decrease with the frequency increase are in contradiction with conventional LK dynamics of polarization reversal. The origin of the paradoxical behavior is the sluggish ion charges dynamics, which screen the polarization at low frequencies ("ferroionic" switching) and strongly retard with respect to polarization reversal at high frequencies ("separated" switching). The retardation appears because LK relaxation time is much smaller than the time of the surface charge relaxation. Hence, the ferroelectric and surface ion states are mainly inseparable in thermodynamic equilibrium and at low frequencies; they become separable in dynamic regime due to the orders of magnitude difference in the relaxation times of ferroelectric polarization and surface ions.

**VIII. DISTINCTIVE FEATURES OF POLARIZATION SCREENING BY SURFACE CHARGES**

We study analytically the surface ionic and ferroelectric states coupling, thermodynamic stability and polarization kinetics in thin ferroelectric films. It has been shown that the electric coupling between surface charges of electrochemical nature and bulk ferroelectricity gives rise to the appearance of coupled ferroionic states in ferroelectric thin films, exquisitely sensitive to external voltage, temperature and pressure. The ferroionic states are mainly inseparable in thermodynamic equilibrium, and become separable in dynamic mode due to the orders of magnitude difference in the relaxation times of ferroelectric polarization and surface charges.

Results of comparative analysis of the distinctive features of polarization screening by ions, by electron conducting electrodes or via the gap presence are summarized in **Table II.**

**TABLE II.** Comparative analysis of the distinctive features of polarization screening

| Physical quantity or property | Screening of polarization by surface charge of ionic nature | Screening of polarization by electron conducting electrodes or via the gap presence |
|---|---|---|
| **Possible stable phases/states** | Non-ferroelectric ionic (**NFE**), ferroionic (**FEI**) and ferroelectric (**FE**) | Paraelectric (**PE**) and ferroelectric (**FE**) |
| **Domain formation** | Single-domain ferroelectric state with out-of-plane polarization is stable if the electric field of the ion layer is high enough | Domain formation is favorable in FE phase at finite extrapolation lengths, as well as under the gap presence |
| **Phase diagram in coordinates temperature $T$-film thickness $h$** | Diffuse boundary between NFE and FEI states essentially depends on the oxygen partial pressure $p_{O2}$. Sharp boundary between FEI and FE states weakly depends on $p_{O2}$ and so the critical thickness can be introduced | Sharp boundary between PE and FE states at the film critical thickness $h_{cr}$ |
| **Constitution diagram in coordinates temperature – pressure** | NFE, FEI and FE states can exist for the film with $h > h_{cr}$. The boundary between NFE and FEI states disappears in the critical point $S_{cr}$. | The film diagram is not dependent on the oxygen partial pressure $p_{O2}$. |
| **Temperature dependence of spontaneous polarization** | Positive and negative polarizations $P_S^+(T)$ and $P_S^-(T)$ are not equivalent for due to the surface charge field. At film thickness $h < h_{cr}$ two branches of the negative spontaneous polarization exists, one of them is unstable. At film thickness $h > h_{cr}$ four branches of the spontaneous polarization exists, two of them are unstable. | At film thickness $h > h_{cr}$ two energetically equivalent states $\pm P_S(T)$ exists. Unstable branches are absent. |
| **Polarization, displacement current and surface charge hysteresis** | Hysteresis loops shape changes and their width decrease with the frequency increase are in contradiction with conventional Landau-Khalatnikov dynamics of polarization reversal. The origin of the paradoxical behavior is the sluggish ion charges, which screen the polarization at low frequencies and strongly retard with respect to polarization reversal at high frequencies. The retardation appears because LK relaxation time is much smaller than the time of the surface charge relaxation. | Conventional Landau-Khalatnikov dynamics of homogeneous (i.e. single-domain) polarization reversal in ferroelectric films for the case of polarization screening by electron conducting electrodes. The gap presence essentially increases the critical thickness of the loop appearance, leads to the loops smearing and causes the phase shift of the displacement current under the frequency increase. |

## ACKNOWLEDGEMENTS

Authors gratefully acknowledge useful discussions with Profs. J.F. Scott, Nava Setter, Karin Rabe, A.K. Tagantsev, D.D. Fong and V.M. Fridkin. This research was sponsored by the Division of Materials Sciences and Engineering, BES, DOE (S.V.K.). Research was conducted at the Center for Nanophase Materials Sciences, which is a DOE Office of Science User Facility.

## SUPPLEMENTARY MATERIALS [99]

### A. Polarization screening by electron conducting electrodes

For comparison with the case of screening by ion charges, we consider a ferroelectric film placed in ideal electric contact with perfect electron conducting electrodes [$\lambda = 0$, see **Fig. 1(b)**], further named as the **"ideal screening"** case. Here, only electrostatic equation (2b) should be solved with the boundary conditions $\varphi_d|_{z=0} = U$ and $\varphi_f|_{z=h} = 0$. For this case, approximate analytical expressions for the dependences of spontaneous polarization and thermodynamic coercive field on the temperature and film thickness have been derived and shown to be valid with a high degree of accuracy (see Ref.[16] from the main text). The expressions are relatively simple for the films with the second order phase transition from ferroelectric (**FE**) to paraelectric (**PE**) phase, for which $a_{333} = 0$. In this case

$$P_3^S = P_S^{bulk}\sqrt{1 - h_{cr}/h}, \qquad E_c = E_c^{bulk}\sqrt{(1 - h_{cr}/h)^3} \qquad (A.1)$$

where $P_S^{bulk} = \sqrt{-a_3/a_{33}}$ and $E_c^{bulk} = 2a_3 P_S^{bulk}/\sqrt{27}$ are temperature-dependent spontaneous polarization and thermodynamic coercive field of a bulk ferroelectric correspondingly. The critical thickness of the film FE-PE phase transition has the form

$$h_{cr}^0(T) \approx \frac{g_{33}}{\alpha_T(T_C - T)}\left(\frac{1}{\Lambda_+ + L_C} + \frac{1}{\Lambda_- + L_C}\right), \qquad (A.2)$$

where $L_C = \sqrt{\varepsilon_0 \varepsilon_{33}^b g_{33}}$ is the longitudinal correlation length.

### B. Polarization screening by electrodes in the presence of dielectric gap

More complex case is the film placed between electron conducting metal electrodes, without any ion screening charges ($\sigma = 0$) at the film surfaces, but in the presence of dielectric gap ($\lambda > 0$) [see **Fig. 1(c)**]. In this case, the critical thickness of the single-domain state instability can be estimated from the condition

$$\alpha + \left(g_{33}\left(\frac{1}{\Lambda_+ + L_C} + \frac{1}{\Lambda_- + L_C}\right) + \frac{\tilde{\lambda}}{\varepsilon_0}\right)\frac{1}{\varepsilon_{33}^b \tilde{\lambda} + h} = 0, \qquad (A.3)$$

where the value $\tilde{\lambda} = \lambda/\varepsilon_d$ plays the role of effective thickness of the dielectric gap (or effective screening radius of electrodes), responsible for imperfect screening.

From the Eq. (A.3) we obtain that

$$h_{SDI}^\lambda(T) = \frac{g_{33}}{\alpha_T(T_C - T)}\left(\frac{1}{\Lambda_+ + L_C} + \frac{1}{\Lambda_- + L_C}\right) + \frac{\lambda}{\varepsilon_d}\left(\frac{1}{\varepsilon_0 \alpha_T(T_C - T)} - \varepsilon_{33}^b\right). \qquad (A.4)$$

Note that $h_{SDI}^\lambda(T) - h_{cr}^0(T) = \frac{\lambda}{\varepsilon_d}\left(\frac{1}{\varepsilon_0 \alpha_T(T_C - T)} - \varepsilon_{33}^b\right)$. Numerically, $h_{SDI}^\lambda(T)$ is usually much higher than $h_{cr}^0(T)$ given by Eq.(A.2). However, one should take into account that under the presence of dielectric gap the film easily splits into ferroelectric domains at temperatures below $T_c$, providing alternative pathway for reduction of electrostatic energy.

## C. Derivation of the coupled equations for the case of surface ion layer presence

Let us consider 1D case, also we suppose that distribution is almost homogeneous $P_3 \approx \langle P_3 \rangle$ (which is valid for $\Lambda_\pm \gg \sqrt{\varepsilon_0 \varepsilon_{33}^b g}$ with the latter length-scale being very small). For the case the solution for potential distribution exists:

$$\varphi_d = U - \frac{z+\lambda}{\lambda}(U-\Psi) \qquad \varphi_f = (h-z)\frac{\Psi}{h} \qquad (A.5)$$

These potentials satisfy the boundary conditions (3a). The constant $\Psi$ should be determined from the condition (3b).

$$\varepsilon_0 \varepsilon_d \frac{U-\Psi}{\lambda} = \langle P_3 \rangle + \varepsilon_0 \varepsilon_{33}^b \frac{\Psi}{h} - \sigma(\Psi) \qquad (A.6)$$

Averaging of equilibrium Eq.(1a) and application of boundary conditions (1b) leads to the equation

$$a_3 \langle P_3 \rangle + a_{33}\langle P_3^3 \rangle + a_{333}\langle P_3^5 \rangle + \frac{g}{h}\left(\left.\frac{P_3}{\Lambda_+}\right|_{z=t+\lambda} + \left.\frac{P_3}{\Lambda_-}\right|_{z=\lambda}\right) = \langle E_3 \rangle \qquad (A.7)$$

That in its turn yields the following approximate equation

$$\left(a_3 + \frac{g}{h}\left(\frac{1}{\Lambda_+} + \frac{1}{\Lambda_-}\right)\right)\langle P_3 \rangle + a_{33}\langle P_3 \rangle^3 + a_{333}\langle P_3 \rangle^5 \approx \frac{\Psi}{h} \qquad (A.8)$$

After elementary transformations the system of nonlinear equations (A.5)-(A.8) for $\langle P_3 \rangle$ and $\Psi$ can be rewritten as:

$$U = \frac{\lambda}{\varepsilon_0 \varepsilon_d}\left(\langle P_3 \rangle + \left(\frac{\varepsilon_0 \varepsilon_d}{\lambda} + \frac{\varepsilon_0 \varepsilon_{33}^b}{h}\right)\Psi - \sigma(\Psi)\right) \qquad (A.9a)$$

$$\left(h + \lambda\frac{\varepsilon_0 \varepsilon_{33}^b}{\varepsilon_0 \varepsilon_d}\right)\left(\left(a_3 + \frac{g}{h}\left(\frac{1}{\Lambda_+} + \frac{1}{\Lambda_-}\right)\right)\langle P_3 \rangle + a_{33}\langle P_3 \rangle^3 + a_{333}\langle P_3 \rangle^5\right) - U +$$
$$+ \frac{\lambda}{\varepsilon_0 \varepsilon_d}\left(\langle P_3 \rangle - \sigma\left(\left[\left(a_3 + \frac{g}{h}\left(\frac{1}{\Lambda_+} + \frac{1}{\Lambda_-}\right)\right)\langle P_3 \rangle + a_{33}\langle P_3 \rangle^3 + a_{333}\langle P_3 \rangle^5\right]h\right)\right) = 0 \qquad (A.9b)$$

From Eqs.(A.9), further self-consistent modeling was based on the following system of non-linear equation that is in fact coupled Eqs.(7) from the main text:

$$\frac{\partial \sigma}{\partial t} + \frac{\sigma}{\tau} = \frac{1}{\tau}\sigma_0 \left[ \frac{U(\varepsilon_d/\lambda) + (\sigma - \langle P_3 \rangle)/\varepsilon_0}{(\varepsilon_d/\lambda) + (\varepsilon_{33}^b/h)} \right] \quad \text{(A.10a)}$$

$$\Gamma \frac{\partial \langle P_3 \rangle}{\partial t} + \left( a_3 + \frac{g}{h}\left( \frac{1}{\Lambda_+} + \frac{1}{\Lambda_-} \right) \right)\langle P_3 \rangle + a_{33}\langle P_3 \rangle^3 + a_{333}\langle P_3 \rangle^5 = \frac{1}{h}\left( \frac{U(\varepsilon_d/\lambda) + (\sigma - \langle P_3 \rangle)/\varepsilon_0}{(\varepsilon_d/\lambda) + (\varepsilon_{33}^b/h)} \right) \quad \text{(A.10b)}$$

## APPENDIX B.

### A. Polarization relaxation to a single-domain state

To illustrate the effect of the internal field created by ions on polarization relaxation to the steady state, **Fig. S1** shows the polarization relaxation to the steady state, namely the distributions of the out-of-plane polarization component $P_3$ in the 100-nm film at successive time moments $t = 0$, $10\tau_K$, $15\tau_K$ and $20\tau_K$ (from left to right). Top and bottom rows are calculated for "symmetric" ($\Delta G_1^{00} = \Delta G_2^{00} = 1\text{eV}$) and "asymmetric" ($\Delta G_1^{00} - \Delta G_2^{00} = 0.9\text{ eV}$) dependences of surface charge $\sigma$ on the potential $\varphi$, respectively. As anticipated the equilibrium state is multi-domain in the case $\Delta G_1^{00} = \Delta G_2^{00}$ (top row in **Fig.S1**). In contrast, up- and down-directed polarizations are not energetically equivalent in the case $\Delta G_1^{00} \neq \Delta G_2^{00}$ because of the poling field (bottom row in **Fig.S1**). The field effective value is nonzero at zero applied voltage, $E_{eff}(U = 0, \sigma) \neq 0$, because the ion charges $\sigma(U)$, which induce the field, have nonzero charge density at $U = 0$ for different ion formation energies $\Delta G_1^{00} \neq \Delta G_2^{00}$ [see **Figs. 2(b)-(c)**]. With time a poly-domain seeding becomes unstable and the film covered by ions relaxes to the single-domain state at $t > 20\tau_K$.

To explain the result shown in **Fig. S1**, we compared the three electrostatic free energies of the thin $Pb_{0.5}Zr_{0.5}TiO_3$ film at zero applied voltage $U = 0$, namely with domain structure but without top ion layer (case 1); with domain structure and with top ion layer that provides effective chemical screening (case 2); without domain structure and with top ion layer that provides effective chemical screening (case 3). For all cases the film is in a perfect electric contact with electron conducting bottom electrode. Appeared the case 3 is the most energetically preferable as anticipated if only the polarizing field is rather high, which is true if the difference between the ion formation energies $\left(\Delta G_1^{00} - \Delta G_2^{00}\right)$ is enough high ($\geq 0.3$ eV).

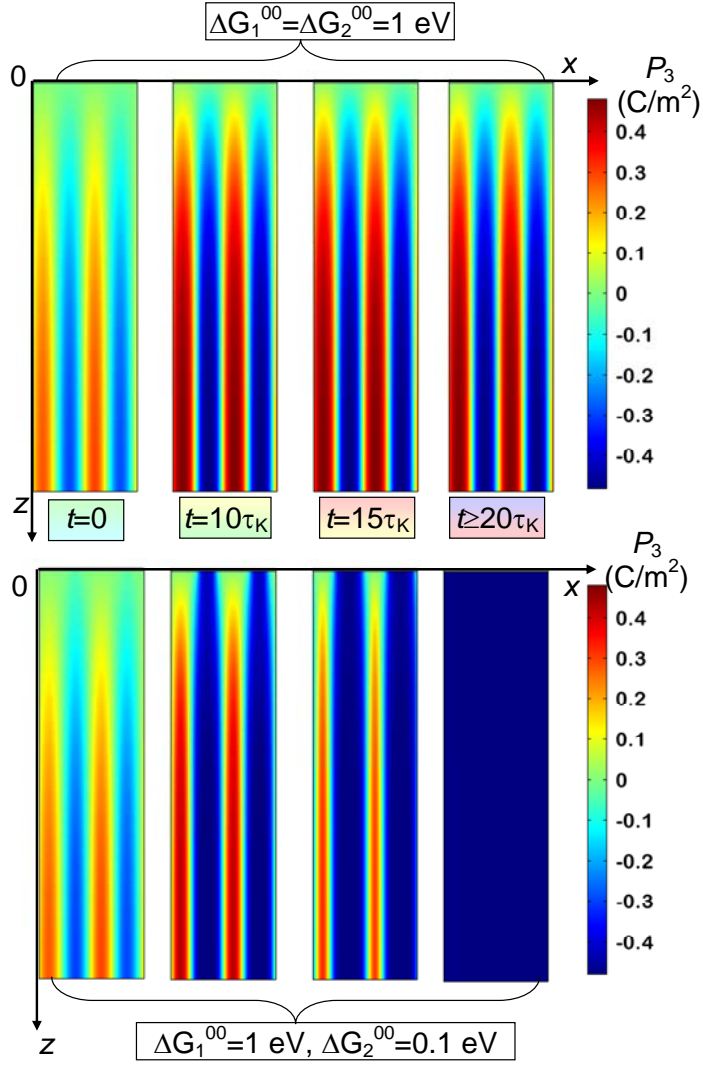

**FIGURE S1.** Color maps of the distribution of the out-of-plane polarization component $P_3$ in the PZT film at successive time moments $t = 0$, $10\tau_K$, $15\tau_K$ and $20\tau_K$ (from left to right). Top and bottom rows are calculated for "symmetric" ($\Delta G_1^{00} = \Delta G_2^{00}$) and "asymmetric" ($\Delta G_1^{00} \neq \Delta G_2^{00}$) dependences of surface charge σ on the potential φ, respectively. External voltage is zero. Film thickness is 100 nm (vertical scale). The scale of the color bar is in $C/m^2$. Two periods of the domain structure are shown.

## APPENDIX C.

### A. Voltage dependences of the effective field and average polarization

Voltage dependences of the effective field and average polarization calculated for ideal screening are shown in **Fig.S2**.

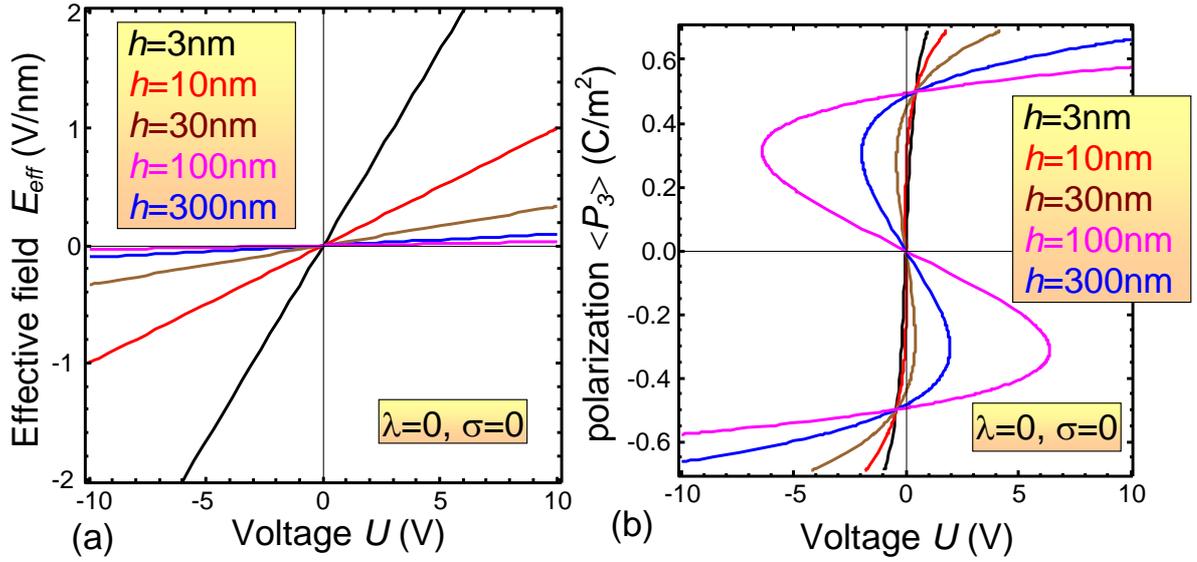

**FIGURE S2**. Dependence of the average electric field $E_{eff}(U)$ in ferroelectric **(a)** and average polarization $\langle P_3 \rangle$ **(b)** on applied voltage $U$ calculated for different thickness of ferroelectric film $h$ = (3, 1000) nm (black, red, brown, purple and blue curves). Parameters T = 300 K and λ = 0. Other parameters used in our calculations are listed in **Table I, main text**.

**B. Temperature dependences of the average polarization at different voltages**

     **Figure S3** shows the temperature dependences of the average polarization partially screened by ion charges calculated for different values of film thickness $h$ (plots **(a)-(d)**). Each curve corresponds to one of the voltage values $U$. The figure clearly shows how these dependencies differ from the ones calculated for σ = 0 and shown in **Fig.S3 (b)-(d)**. In fact, analyzing the significant differences we come to the conclusion that the impact of such screening on the polarization is not trivial and can not be reduced to the decrease of dielectric gap thickness or to the change of the dielectric parameters. The reason of this is that the voltage dependence σ(U) is essentially nonlinear in the overpotential Ψ according to Eq. (10), while reducing the thickness λ of the gap does not introduce any additional nonlinearity in the system.

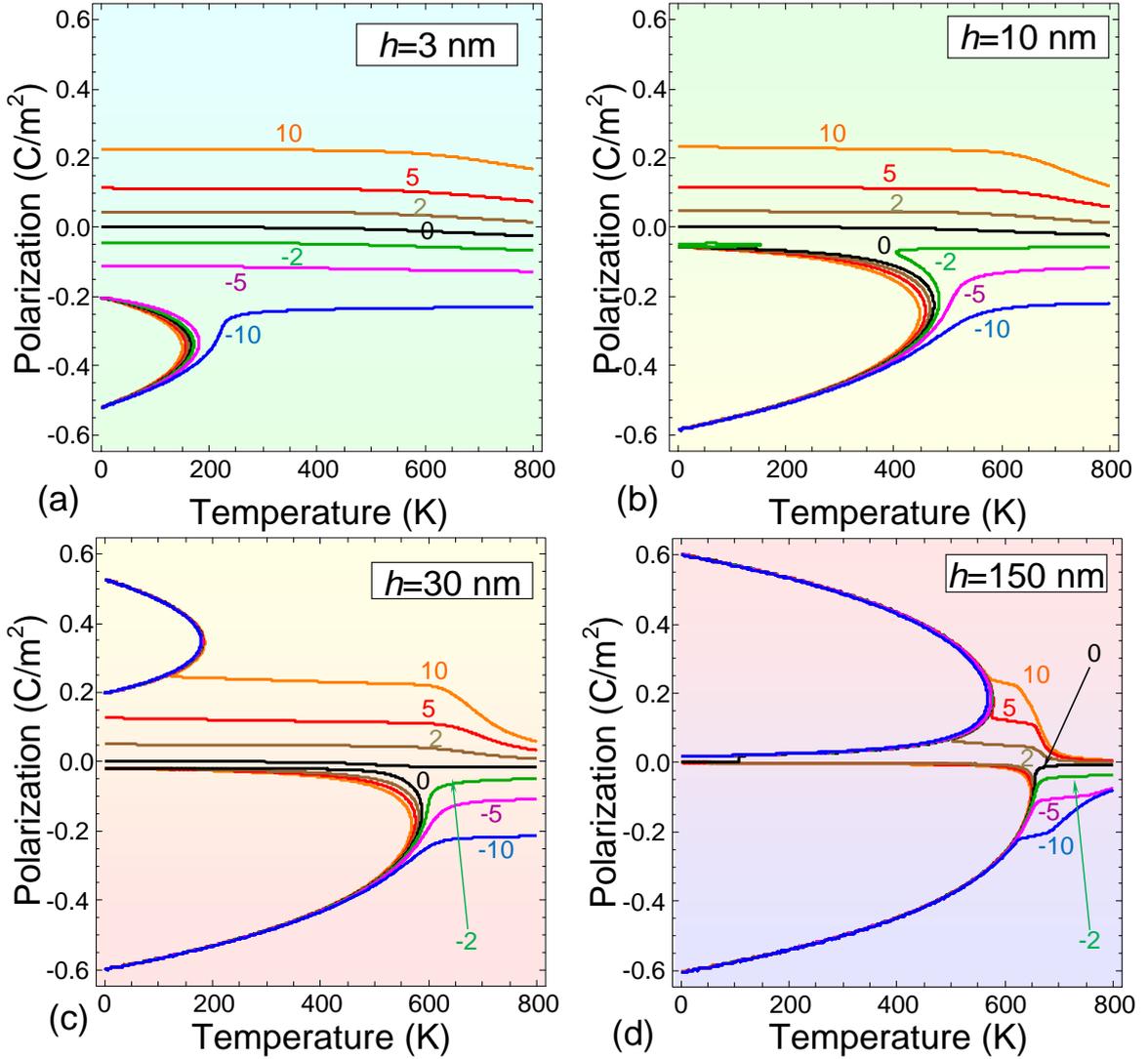

**FIGURE S3.** Temperature dependences of the average polarization calculated in the case of its screening by surface ion charges for different values of film thickness $h$ = 3, 10, 30 and 150 nm (plots **(a), (b), (c)** and **(d)** respectively). Each curve corresponds to one of the applied voltage values $U = -10, -5, -2, 0, 2, 5, 10$ V (numbers near the curves of different colors). Parameters $p_{O2} = 10^{-1}$ bar, $A_1 = A_2 = 2\times 10^{-19}$ m$^{-2}$, $\Delta G_1^{00} = 1$ eV and $\Delta G_2^{00} = 0.1$ eV.

**C. Polarization reversal in the case of polarization screening by electron conducting electrodes.**

To analyze and compare the **case 1** and **3** of dynamic polarization screening by electrodes without surface ions, the time dependences of $\langle P_3 \rangle$ and $\partial \langle P_3 \rangle / \partial t$ were calculated for several values of the film thickness $h = (3\text{-}300)$ nm for $\lambda = 0$ [**Fig.S4**], and the two values of dimensionless frequency $\omega\tau_K = 10^{-2}$ [parts **(a, c)**] and $\omega\tau_K = 0.3$ [parts **(b, d)**], respectively. Very narrow loops correspond to the thinnest film in a paraelectric phase at low frequency $\omega\tau_K = 10^{-2}$ [see the black loop for 3-nm film in **Fig. S4 (a)**]. Then the loop width gradually increases with the film thickness increase (red, purple and blue loops in **Fig. S4(a)**), and the loop shape tends to the square one corresponding to the ferroelectric film of thickness 300 nm for λ = 0 (blue loop in **Figs. S4 (a)**). The maxima of displacement currents $\partial \langle P_3 \rangle / \partial t$ correspond to the change of polarization sign at coercive voltage (see **Fig. S4(c)**). The maxima of $\partial \langle P_3 \rangle / \partial t$ are very sharp and high for λ = 0.

The shape and width of the polarization reversal loops in **Figs.S4 (b)** changes significantly with increasing the frequency in thirty times ($\omega\tau_K = 0.3$). They become essentially smoother and wider; and the loop with finite width occurs for the thinnest films. Corresponding current maxima become much lower and wider (compare different loops in **Figs.S4 (c) and (d)**). At very high frequencies $\omega\tau_K > 1$ (not shown in the figures) all the loops become strongly "inflated".

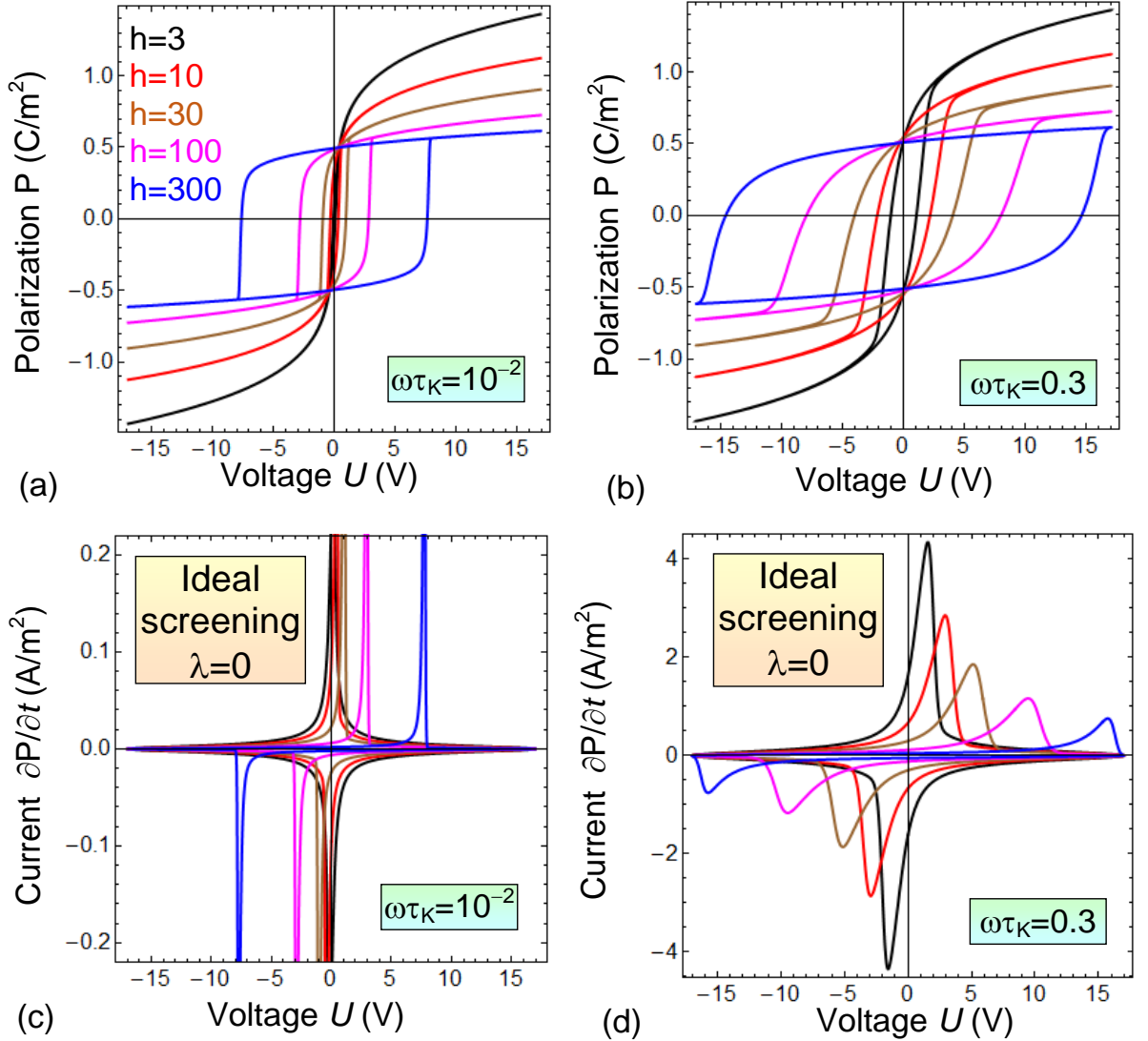

**FIGURE S4**. **Polarization reversal in the case of polarization screening by electron conducting electrodes.** Average polarization $\langle P_3 \rangle$ **(a,b)** and displacement current $\partial \langle P_3 \rangle / \partial t$ **(c,d)** dependences on the amplitude $U$ of applied periodic voltage calculated for its several frequencies ω, $\omega\tau_K$ = $10^{-2}$ **(a, c)** and 0.3 **(b, d)**,. Different colors of the curves correspond to different thickness of the film $h$ = 3 nm (black curves), 10 nm (red curves), 30 nm (brown curves), 100 (purple curves) and 300 nm (blue d curves). Temperature T = 300 K, λ = 0, the frequency is in the units of Landau-Khalatnikov time $\tau_K$. Other parameters used in our calculations are listed in **Table I**.

The conclusions made from **Fig.S4** are in a complete agreement with conventional Landau-Khalatnikov dynamics of homogeneous (i.e. single-domain) polarization reversal in ferroelectric films described by time-dependent LGD equation.